\documentclass[11pt,draftcls,onecolumn]{IEEEtran} 
\usepackage{epsfig,graphics,color,subfigure,wrapfig,hyperref,url}
\usepackage{amssymb,amsmath,algorithm,times}
\usepackage{slashbox}

\begin{document}
\title{Regularized Modified BPDN for Noisy Sparse Reconstruction with Partial Erroneous Support and Signal Value Knowledge
\thanks{Copyright (c) 2011 IEEE. Personal use of this material is permitted. However, permission to use this material for any other purposes must be obtained from the IEEE by sending a request to pubs-permissions@ieee.org.}
\thanks{A part of this work was presented at IEEE International Conference on Acoustics,speech and signal processing (ICASSP), 2010 \cite{modBPDN}. This research was partially supported by NSF grants ECCS-0725849 and CCF-0917015. The authors are with the Department of Electrical and Computer Engineering, Iowa State University, Ames,
IA 50010 USA (Email: luwei@iastate.edu; namrata@iastate.edu. Phone: 515-294-4012. Fax: 515-294-8432.)  }  }
\author{Wei Lu and Namrata Vaswani}

\maketitle

\newcommand{\Dnum}{D_{num}}
\newcommand{\pss}{p^{**,i}}
\newcommand{\fr}{f_{r}^i}

\newcommand{\A}{{\cal A}}
\newcommand{\Z}{{\cal Z}}
\newcommand{\B}{{\cal B}}
\newcommand{\R}{{\cal R}}
\newcommand{\reg}{{\cal G}}
\newcommand{\const}{\mbox{const}}

\newcommand{\trace}{\mbox{trace}}

\newcommand{\hsim}{{\hspace{0.0cm} \sim  \hspace{0.0cm}}}
\newcommand{\he}{{\hspace{0.0cm} =  \hspace{0.0cm}}}

\newcommand{\vect}[2]{\left[\begin{array}{cccccc}
     #1 \\
     #2
   \end{array}
  \right]
  }

\newcommand{\matr}[2]{ \left[\begin{array}{cc}
     #1 \\
     #2
   \end{array}
  \right]
  }
\newcommand{\vc}[2]{\left[\begin{array}{c}
     #1 \\
     #2
   \end{array}
  \right]
  }

\newcommand{\gdot}{\dot{g}}
\newcommand{\Cdot}{\dot{C}}
\newcommand{\re}{\mathbb{R}}
\newcommand{\n}{{\cal N}}  
\newcommand{\N}{{\overrightarrow{\bf N}}}  
\newcommand{\chat}{\tilde{C}_t}
\newcommand{\chati}{\chat^i}

\newcommand{\cmin}{C^*_{min}}
\newcommand{\twi}{\tilde{w}_t^{(i)}}
\newcommand{\twj}{\tilde{w}_t^{(j)}}
\newcommand{\wi}{{w}_t^{(i)}}
\newcommand{\twio}{\tilde{w}_{t-1}^{(i)}}

\newcommand{\tWi}{\tilde{W}_n^{(m)}}
\newcommand{\tWj}{\tilde{W}_n^{(k)}}
\newcommand{\Wi}{{W}_n^{(m)}}
\newcommand{\tWio}{\tilde{W}_{n-1}^{(m)}}

\newcommand{\ds}{\displaystyle}

\newcommand{\SAR}{S$\!$A$\!$R }
\newcommand{\MAR}{MAR}
\newcommand{\MMRF}{MMRF}
\newcommand{\AR}{A$\!$R }
\newcommand{\GMRF}{G$\!$M$\!$R$\!$F }
\newcommand{\DTM}{D$\!$T$\!$M }
\newcommand{\MSE}{M$\!$S$\!$E }
\newcommand{\RCS}{R$\!$C$\!$S }
\newcommand{\uomega}{\underline{\omega}}
\newcommand{\y}{v}
\newcommand{\x}{w}
\newcommand{\lu}{\mu}
\newcommand{\g}{g}
\newcommand{\s}{{\bf s}}
\newcommand{\bft}{{\bf t}}
\newcommand{\refmap}{{\cal R}}
\newcommand{\totrefl}{{\cal E}}
\newcommand{\beq}{\begin{equation}}
\newcommand{\eeq}{\end{equation}}
\newcommand{\bdm}{\begin{displaymath}}
\newcommand{\edm}{\end{displaymath}}
\newcommand{\hatz}{\hat{z}}
\newcommand{\hatu}{\hat{u}}
\newcommand{\tilz}{\tilde{z}}
\newcommand{\tilu}{\tilde{u}}
\newcommand{\hhatz}{\hat{\hat{z}}}
\newcommand{\hhatu}{\hat{\hat{u}}}
\newcommand{\tilc}{\tilde{C}}
\newcommand{\hatc}{\hat{C}}
\newcommand{\tim}{n}

\newcommand{\ssp}{\renewcommand{\baselinestretch}{1.0}}
\newcommand{\defd}{\mbox{$\stackrel{\mbox{$\triangle$}}{=}$}}
\newcommand{\goes}{\rightarrow}
\newcommand{\tends}{\rightarrow}
\newcommand{\defn}{\triangleq} 
\newcommand{\se}{&=&}
\newcommand{\sdefn}{& \defn  &}
\newcommand{\sle}{& \le &}
\newcommand{\sge}{& \ge &}
\newcommand{\plusminus}{\stackrel{+}{-}}
\newcommand{\Ey}{E_{Y_{1:t}}}
\newcommand{\ey}{E_{Y_{1:t}}}

\newcommand{\equivto}{\mbox{~~~which is equivalent to~~~}}
\newcommand{\nonzero}{i:\pi^n(x^{(i)})>0}
\newcommand{\nonzeroc}{i:c(x^{(i)})>0}

\newcommand{\supn}{\sup_{\phi:||\phi||_\infty \le 1}}
\newtheorem{theorem}{Theorem}
\newtheorem{lemma}{Lemma}
\newtheorem{corollary}{Corollary}
\newtheorem{definition}{Definition}
\newtheorem{remark}{Remark}
\newtheorem{example}{Example}
\newtheorem{ass}{Assumption}
\newtheorem{fact}{Fact}
\newtheorem{heuristic}{Heuristic}
\newcommand{\eps}{\epsilon}
\newcommand{\bd}{\begin{definition}}
\newcommand{\ed}{\end{definition}}
\newcommand{\udq}{\underline{D_Q}}
\newcommand{\td}{\tilde{D}}
\newcommand{\epsinv}{\epsilon_{inv}}
\newcommand{\al}{\mathcal{A}}

\newcommand{\bfx} {\bf X}
\newcommand{\bfy} {\bf Y}
\newcommand{\bfz} {\bf Z}
\newcommand{\ddas}{\mbox{${d_1}^2({\bf X})$}}
\newcommand{\ddbs}{\mbox{${d_2}^2({\bfx})$}}
\newcommand{\dda}{\mbox{$d_1(\bfx)$}}
\newcommand{\ddb}{\mbox{$d_2(\bfx)$}}
\newcommand{\xinc}{{\bfx} \in \mbox{$C_1$}}
\newcommand{\eqa}{\stackrel{(a)}{=}}
\newcommand{\eqb}{\stackrel{(b)}{=}}
\newcommand{\eqe}{\stackrel{(e)}{=}}
\newcommand{\leqc}{\stackrel{(c)}{\le}}
\newcommand{\leqd}{\stackrel{(d)}{\le}}

\newcommand{\leqa}{\stackrel{(a)}{\le}}
\newcommand{\leqb}{\stackrel{(b)}{\le}}
\newcommand{\leqe}{\stackrel{(e)}{\le}}
\newcommand{\leqf}{\stackrel{(f)}{\le}}
\newcommand{\leqg}{\stackrel{(g)}{\le}}
\newcommand{\leqh}{\stackrel{(h)}{\le}}
\newcommand{\leqi}{\stackrel{(i)}{\le}}
\newcommand{\leqj}{\stackrel{(j)}{\le}}

\newcommand{\w}{{W^{LDA}}}
\newcommand{\halpha}{\hat{\alpha}}
\newcommand{\hsigma}{\hat{\sigma}}
\newcommand{\slmax}{\sqrt{\lambda_{max}}}
\newcommand{\slmin}{\sqrt{\lambda_{min}}}
\newcommand{\lmax}{\lambda_{max}}
\newcommand{\lmin}{\lambda_{min}}

\newcommand{\da} {\frac{\alpha}{\sigma}}
\newcommand{\chka} {\frac{\check{\alpha}}{\check{\sigma}}}
\newcommand{\sumo}{\sum _{\underline{\omega} \in \Omega}}
\newcommand{\distance}{d\{(\hatz _x, \hatz _y),(\tilz _x, \tilz _y)\}}
\newcommand{\col}{{\rm col}}
\newcommand{\rcs}{\sigma_0}
\newcommand{\CalR}{{\cal R}}
\newcommand{\df}{{\delta p}}
\newcommand{\dq}{{\delta q}}
\newcommand{\dZ}{{\delta Z}}
\newcommand{\pprime}{{\prime\prime}}

\newcommand{\vn}{N}

\newcommand{\bv}{\begin{vugraph}}
\newcommand{\ev}{\end{vugraph}}
\newcommand{\bi}{\begin{itemize}}
\newcommand{\ei}{\end{itemize}}
\newcommand{\ben}{\begin{enumerate}}
\newcommand{\een}{\end{enumerate}}
\newcommand{\be}{\protect\[}
\newcommand{\ee}{\protect\]}
\newcommand{\bean}{\begin{eqnarray*} }
\newcommand{\eean}{\end{eqnarray*} }
\newcommand{\bea}{\begin{eqnarray} }
\newcommand{\eea}{\end{eqnarray} }
\newcommand{\nn}{\nonumber}
\newcommand{\ba}{\begin{array} }
\newcommand{\ea}{\end{array} }
\newcommand{\ep}{\mbox{\boldmath $\epsilon$}}
\newcommand{\epp}{\mbox{\boldmath $\epsilon '$}}
\newcommand{\Lep}{\mbox{\LARGE $\epsilon_2$}}
\newcommand{\und}{\underline}
\newcommand{\pdif}[2]{\frac{\partial #1}{\partial #2}}
\newcommand{\odif}[2]{\frac{d #1}{d #2}}
\newcommand{\dt}[1]{\pdif{#1}{t}}
\newcommand{\urho}{\underline{\rho}}

\newcommand{\spc}{{\cal S}}
\newcommand{\tspc}{{\cal TS}}

\newcommand{\uv}{\underline{v}}
\newcommand{\us}{\underline{s}}
\newcommand{\uc}{\underline{c}}
\newcommand{\utheta}{\underline{\theta}^*}
\newcommand{\ualpha}{\underline{\alpha^*}}

\newcommand{\uxy}{\underline{x}^*}
\newcommand{\uxyj}{[x^{*}_j,y^{*}_j]}
\newcommand{\arcl}[1]{arclen(#1)}
\newcommand{\one}{{\mathbf{1}}}

\newcommand{\uxyjt}{\uxy_{j,t}}
\newcommand{\E}{\mathbb{E}}

\newcommand{\rhomat}{\left[\begin{array}{c}
                        \rho_3 \ \rho_4 \\
                        \rho_5 \ \rho_6
                        \end{array}
                   \right]}
\newcommand{\deltat}{\tau} 
\newcommand{\deltatt}{\Delta t_1}
\newcommand{\ceil}[1]{\ulcorner #1 \urcorner}

\newcommand{\xxi}{x^{(i)}}
\newcommand{\txi}{\tilde{x}^{(i)}}
\newcommand{\txj}{\tilde{x}^{(j)}}

\newcommand{\mi}[1]{{#1}^{(m,i)}}

\setlength{\arraycolsep}{0.03cm}
\newcommand{\tDelta}{{\tilde{\Delta}}}
\newcommand{\xhat}{\hat{x}}
\newcommand{\Nhat}{\hat{N}}
\newtheorem{assum}{Assumption}
\newtheorem{proposition}{Proposition}

\begin{abstract}
We study the problem of sparse reconstruction from noisy undersampled measurements when the following knowledge is available. (1) We are given partial, and partly erroneous, knowledge of the signal's support, denoted by $T$. (2) We are also given an erroneous estimate of the signal values on $T$, denoted by $(\hat{\mu})_T$.
In practice, both of these may be available from prior knowledge. Alternatively, in recursive reconstruction applications, like real-time dynamic MRI, one can use the support estimate and the signal value estimate from the previous time instant as $T$ and $(\hat{\mu})_T$.
In this work, we introduce regularized modified-BPDN (reg-mod-BPDN) to solve this problem and obtain computable bounds on its reconstruction error. Reg-mod-BPDN tries to find the signal that is sparsest outside the set $T$, while being ``close enough" to $(\hat{\mu})_T$ on $T$ and while satisfying the data constraint. Corresponding results for modified-BPDN and BPDN follow as direct corollaries. A second key contribution is an approach to obtain computable error bounds that hold without any sufficient conditions. This makes it easy to compare the bounds for the various approaches. Empirical reconstruction error comparisons with many existing approaches are also provided.
\end{abstract}

\begin{keywords}
compressive sensing, sparse reconstruction, modified-CS, partially known support
\end{keywords}
\section{Introduction}

The goal of this work is to solve the sparse recovery problem \cite{BPDN,justrelax,candescs,Donoho,rice}. We try to reconstruct an $m$-length sparse vector, $x$, with support, $N$, from an $n< m$ length noisy measurement vector, $y$, satisfying
\begin{equation}
y \triangleq Ax+w
\label{obsmod}
\end{equation}
when the following two things are available: (i) partial, and partly erroneous, knowledge of the signal's support, denoted by $T$; and (ii) an erroneous estimate of the signal values on $T$, denoted by $(\hat{\mu})_T$. In (\ref{obsmod}), $w$ is the measurement noise and $A$ is the measurement matrix.
For simplicity, in this work, we just refer to {\em $x$ as the signal} and to {\em $A$ as the measurement matrix}.
However, in general, $x$ is the sparsity basis vector (which is either the signal itself or some linear transform of the signal) and $A = H \Phi$ where $H$ is the measurement matrix and $\Phi$ is the sparsity basis matrix. If $\Phi$ is the identity matrix then $x$ is the signal itself.%

The true support of the signal, $N$, can be rewritten as
\begin{equation}
N = T \cup \Delta \setminus \Delta_e
\end{equation}
where
\begin{equation}
\Delta \triangleq N \setminus T \ \text{and} \ \Delta_e \triangleq T \setminus N
\label{defdelta}
\end{equation}
are the errors in the support estimate, $T^c$ is the complement set of $T$ and $ \setminus $ is the set difference notation ($N \setminus T \triangleq N \cap T^c $).

\newcommand{\muhat}{\hat\mu}

The signal estimate is assumed to be zero along $T^c$, i.e.
\begin{equation}
\hat{\mu} = \left[
              \begin{array}{c}
                (\hat{\mu})_T \\
                \mathbf{0}_{T^c} \\
              \end{array}
            \right]
\label{sigestim}
\end{equation}
and the signal itself can be rewritten as
\begin{eqnarray}
(x)_{N \cup T} &=& (\hat\mu)_{N \cup T} + e \nn \\
\label{signalesterror}
(x)_{N^c}  &=& 0
\end{eqnarray}
where $e$ denotes the error in the prior signal estimate. It is assumed that the error energy, $\|e\|_2^2$, is small compared to the signal energy, $\|x\|_2^2$.

\begin{figure*}
\centerline{
\subfigure[]{
\begin{tabular}{cc}
\epsfig{file = 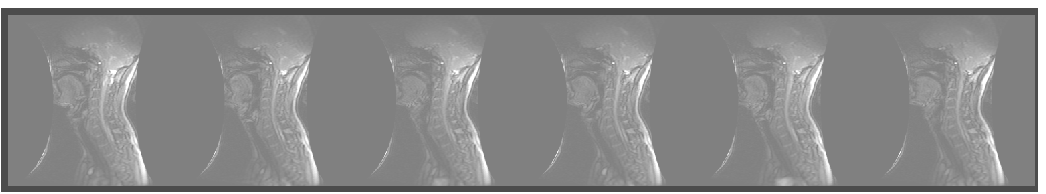,height=1.7cm}   &
\epsfig{file = 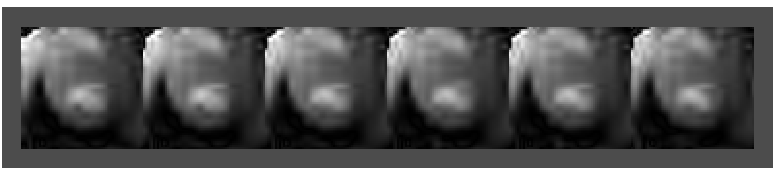, height=1.7cm} \\
{\small (i) a larynx (vocal tract) image sequence} & {\small (ii) cardiac image sequence}
\end{tabular}
}
}
\centerline{
\subfigure[]{
\begin{tabular}{ccc}
\epsfig{file = 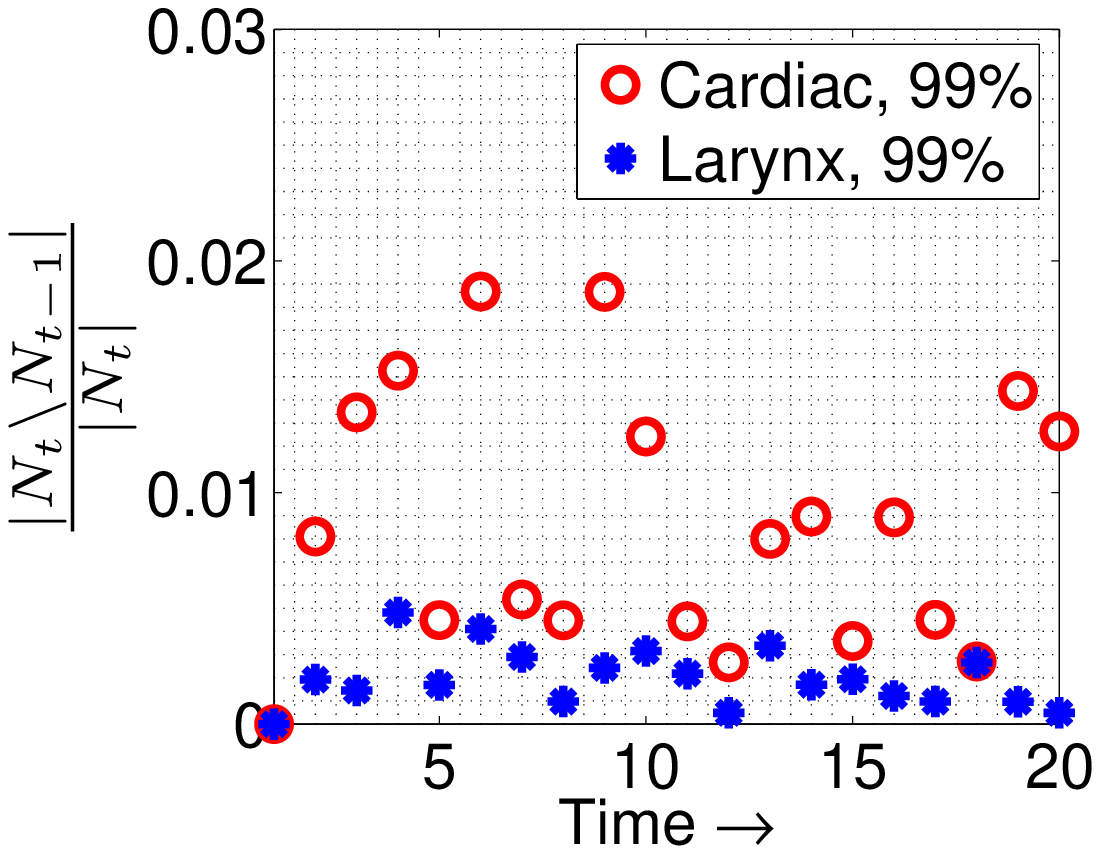,height=3.5cm, width = 6.2cm} & 
\epsfig{file = 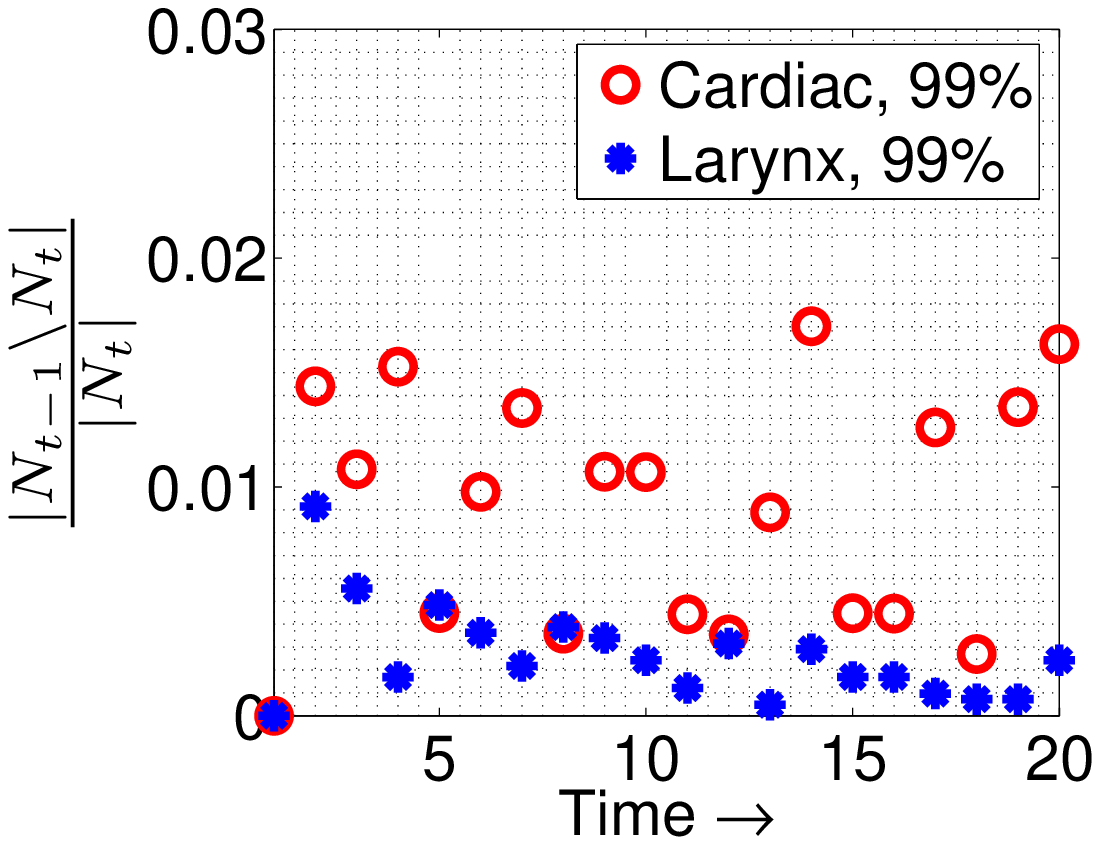,height=3.5cm, width = 6.2cm} &
\epsfig{file = 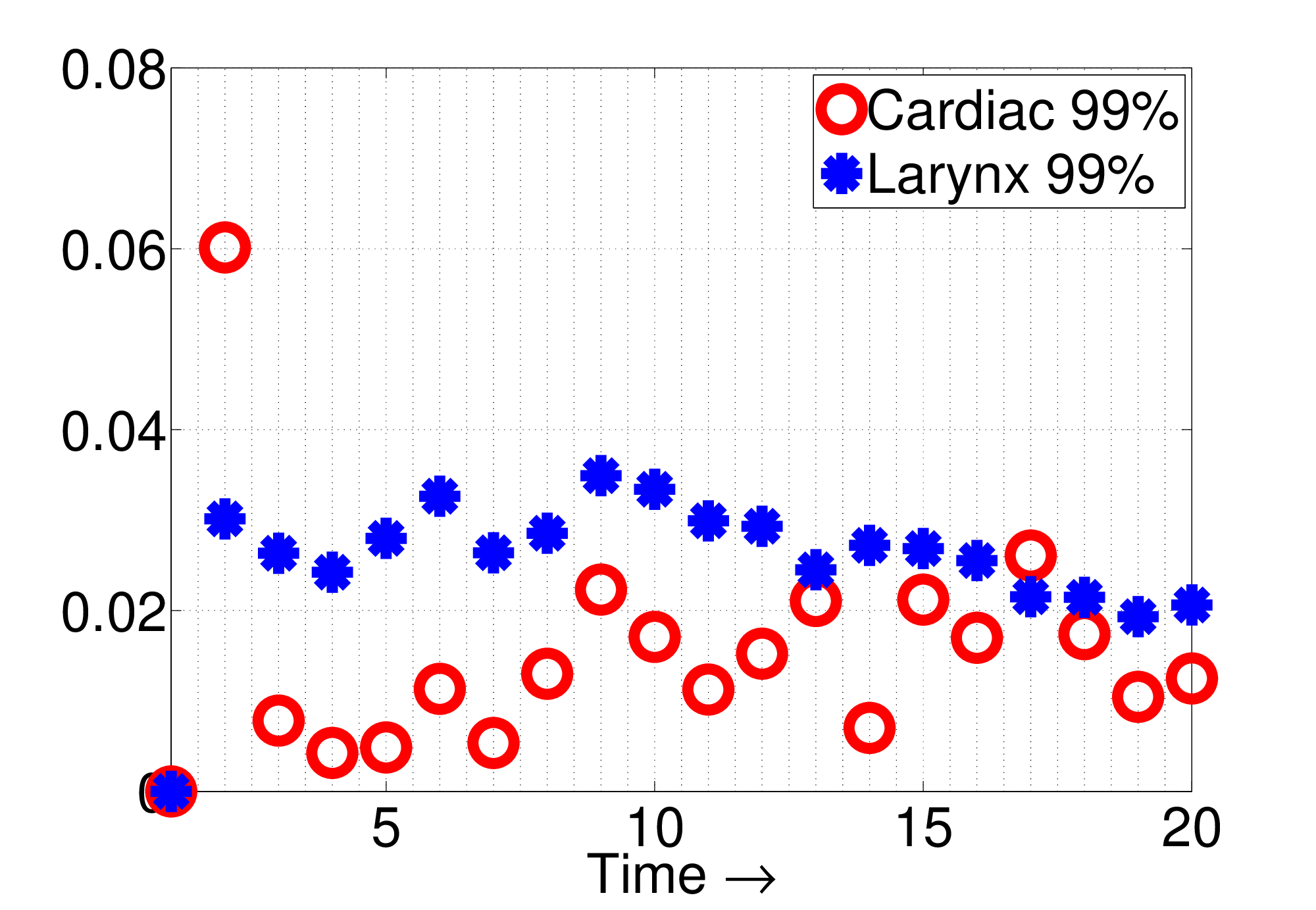,height=3.5cm, width = 6.7cm}\\
{\small (i) support additions, $\frac{|N_t \setminus N_{t-1}|}{|N_t|}$} &  {\small (ii)  support removals,  $\frac{|N_{t-1} \setminus N_{t}|}{|N_t|}$} & {\small  (iii)  signal value change, $\frac{\|(x_t-x_{t-1})_{N_t\cup N_{t-1}}\|_2}{\|(x_t)_{N_t}\|_2}$}
\end{tabular}
}
}
\vspace{-0.1in}
\caption{{\small{In (a), we show two medical image sequences (a cardiac and a larynx sequence).
In (b), $x_t$ is the two-level Daubechies-4 2D discrete wavelet transform (DWT) of the cardiac or the larynx image at time $t$ and the set $N_t$ is its 99\% energy support (the smallest set containing 99\% of the vector's energy). Its size, $|N_t|$ varied between 4121-4183 ($\approx 0.07m$) for larynx and between 1108-1127 ($\approx 0.06m$) for cardiac. {\em Notice that all support changes are less than 2\% of the support size and almost all signal values changes are less than 4\% of $\|(x_t)_{N_t}\|_2$.}}}}
\label{slowchange}
\end{figure*}

In practical applications, $T$ and $\muhat$ may be available from prior knowledge. Alternatively, in applications requiring recursive reconstruction of (approximately) sparse signal or image sequences, with slow time-varying sparsity patterns and slow changing signal values, one can use the support estimate and  the signal value estimate from the previous time instant as the ``prior knowledge". A key domain where this problem occurs is in fast (recursive) dynamic MRI reconstruction from highly undersampled measurements. In MRI, we typically assume that the images are wavelet sparse. 
We show slow support and signal value change for two medical image sequences in Fig. \ref{slowchange}.  From the figure, we can see that the maximum support changes for both sequences are less than 2\% of the support size and almost all signal values' changes are less than $0.16\%$ of the signal energy. Slow signal value change also implies that a signal value is small before it gets removed from the support.
{ Other potential applications include single-pixel camera based real-time video imaging \cite{singlepixelvideo}; video compression; ReProCS (recursive projected CS) based video denoising or video layering (separating video in foreground and background layers) \cite{rrpcp_arxiv,rrpcp_isit}; and spectral domain optical coherence tomography \cite{sd_oct_kang} based dynamic imaging.}


This work has the following contributions.
\ben
\item We introduce regularized modified-BPDN (reg-mod-BPDN) and obtain a computable bound on its reconstruction error using an approach motivated by \cite{justrelax}. Reg-mod-BPDN solves 
\begin{equation}\label{regmodBPDN0}
\min_b \ \  \gamma\|b_{T^c}\|_1  +\frac{1}{2}\|y-Ab\|_2^2 +  \frac{1}{2}\lambda\|b_T-\hat{\mu}_T\|_2^2
\end{equation}
i.e. it tries to find the signal that is sparsest outside the set $T$, while being ``close enough" to $\hat{\mu}_T$ on $T$, and while satisfying the data constraint. Reg-mod-BPDN uses the fact that $T$ is a good estimate of the true support, $N$, and that $\muhat_T$ is a good estimate of $x_T$. In particular, for $i \in \Delta_e$, this implies that  $|\muhat_i|$ is close to zero (since $x_i = 0$ for $i \in \Delta_e$).
\item  Our second key contribution is to show how to use the reconstruction error bound result to obtain another computable bound that holds without any sufficient conditions and is tighter. This allows easy bound comparisons of the various approaches. A similar result for mod-BPDN and BPDN follows as a direct corollary.
\item Reconstruction error comparisons with these and many other existing approaches are also shown.%
\een

\subsection{Notations and Problem Definition}
For any set $T$ and vector $b$, $b_T$ denotes a sub-vector
containing the elements of $b$ with indices in $T$.
$\|b\|_k$ refers to the $\ell_k$ norm of the vector $b$. Also, $\|b\|_0$ counts the number of nonzero elements of $b$.

The notation $T^c$ denotes the set complement of $T$, i.e., $ T^c = \{i \in [1,...,m], i \notin T \}$. $\emptyset$ is the empty set.

We use $'$ for transpose. For the matrix $A$, $A_T$ denotes the sub-matrix containing the columns of $A$ with indices in $T$. The matrix norm $\|A\|_{p}$, is defined as $\|A\|_{p} \triangleq \max_{x\neq 0}\frac{\|Ax\|_p}{\|x\|_p}$.  $I_T$ is an identity matrix on the set of rows and columns indexed by elements in $T$. $\mathbf{0}_{T,S}$ is a zero matrix on the set of rows and columns indexed by elements in $T$ and $S$ respectively.%

The notation $\nabla L(b)$ denotes the gradient of the function $L(b)$ with respect to $b$.

When we say {\em b is supported on $T \cup S$} we mean that the support of $b$ (set of indices where $b$ is nonzero) is a subset of $T \cup S$.

Our goal is to reconstruct a sparse vector, $x$, with support, $N$, from the noisy measurement vector, $y$ satisfying (\ref{obsmod}). We assume partial knowledge of the support, denoted by $T$, and of the signal estimate on $T$, denoted by $(\hat{\mu})_T$. The support estimate may contain errors -- misses, $\Delta$, and extras, $\Delta_e$, defined in (\ref{defdelta}).  The signal estimate, $\muhat$, is assumed to be zero along $T^c$, i.e it satisfies (\ref{sigestim}) and the signal, $x$, satisfies (\ref{signalesterror}).

\subsection{Related Work}
The sparse reconstruction problem, without using any support or signal value knowledge, has been studied for a long
time \cite{BPDN,justrelax,candescs,Donoho,rice}. It tries to find the sparsest signal among all signals that satisfy
the data constraint, i.e. it solves $\min_{b} \|b\|_0 \ s.t. \ y = A\beta$. This brute-force search has exponential complexity. One class of practical approaches to solve this is basis pursuit which replaces $\|b\|_0$ by $\|b\|_1$ \cite{BPDN}. The $\ell_1$ norm is the closest norm to $\ell_0$ that makes the problem convex.
For noisy measurements, the data constraint becomes an inequality constraint. However, this assumes that the noise is bounded and the noise bound is available. In practical applications where this may not be available, one can use the Lagrangian version which solves%
\begin{equation}\label{BPDN}
\min_b \  \gamma\|b\|_1 + \frac{1}{2}\|y-Ab\|_2^2
\end{equation}
This is called {\em basis pursuit denoising (BPDN) \cite{BPDN}}. Since this solves an unconstrained optimization problem, it is also faster. An error bound of BPDN was obtained in \cite{justrelax}. Error bounds for its constrained version were obtained in \cite{candes_rip, candesnoise}.

The problem of sparse reconstruction with partial support knowledge was introduced in our work \cite{modcsisit,modcsjournal};
 and also in parallel in Khajehnejad et al \cite{weightedl1} and in vonBorries et al \cite{camsap07}.
 In \cite{modcsisit,modcsjournal}, we proposed an approach called {\em modified-CS} which tries to find the signal that is
 sparsest outside the set $T$ and satisfies the data constraint. We obtained exact reconstruction conditions for
 it by using the restricted isometry approach \cite{decodinglp}. When measurements are noisy, for the same reasons as above, one can use the Lagrangian version: 
\begin{equation}\label{mb}
\min_b \ \ \gamma\|b_{T^c}\|_1 + \frac{1}{2}\|y-Ab\|_2^2
\end{equation}
We call this {\em modified-BPDN (mod-BPDN)}. Its error was bounded in the conference version of this work \cite{modBPDN}, while the error of its constrained version was bounded in Jacques \cite{iBPDN}.
%

In \cite{weightedl1}, Khajehnejad et al assumed a probabilistic support prior and proposed a {weighted $\ell_1$} solution. They also obtained exact reconstruction thresholds for weighted $\ell_1$ by using the overall approach of Donoho \cite{Dohohowl1}. In Fig. \ref{reconfig}, we show comparisons with the noisy Lagrangian version of {\em weighted $\ell_1$} which solves:
\begin{equation}
\min_b \ \  \gamma\|b_{T^c}\|_1+\gamma'\|b_T\|_1 + \frac{1}{2}\|y-Ab\|_2^2 \label{weightedl1}
\end{equation}

Our earlier work on Least Squares CS-residual (LS-CS) and Kalman Filtered CS-residual (KF-CS) \cite{LSCS,kfcsicip} can also be interpreted as a possible solution for the current problem, although it was proposed in the context of recursive reconstruction of sparse signal sequences. 

Reg-mod-BPDN may also be interpreted as a Bayesian CS or a model-based CS approach. Recent work in this area includes \cite{modelCS, eldar_TIT_2009, bayesianCS, schniter,schniter_hmmtree, TOMP_do, TOMP_bniuk}.

\subsection{Some Related Approaches**}
Before going further, we discuss below {\em a few approaches that are related to, but different from reg-mod-BPDN, and we argue when and why these will be worse than reg-mod-BPDN.} This section may be skipped on a quick reading. We show comparisons with all these in Fig. \ref{reconfig}.

The first is what can be called {\em CS-residual or CS-diff} which computes
\begin{eqnarray}
&& \hat{x} = \hat{\mu} + \hat{b}, \ \text{where $\hat{b}$ solves} \nn \\
&&  \min_{b} \ \ \gamma \|b\|_1 + \frac{1}{2}\|y-A\hat{\mu}-Ab\|_2^2
\label{CSresidual}
\end{eqnarray}
This has the following limitation. It does not use the fact that when $T$ is an accurate estimate of the true support, $(x)_{T^c}$ is much more sparse compared with the full $(x-\hat{\mu})$ {(the support size of $x_{T^c}$ is $|\Delta|$ while that of $(x-\hat{\mu})$ is $|T|+|\Delta|$ which is much larger)}. The exception is if the signal value prior is so strong that $(x-\hat{\mu})$ is zero (or very small) on all or a part of $T$.

CS-residual is also related to LS-CS and KF-CS. LS-CS solves (\ref{CSresidual}) but with $\hat{\mu}_T$ being the LS estimate computed assuming that the signal is supported on $T$ and with $(\muhat)_{T^c}=0$. For a static problem, KF-CS can be interpreted as computing the regularized LS estimate on $T$ and using that as $\hat{\mu}_T$. LS-CS and KF-CS also have a limitation similar to CS-residual.

Another seemingly related approach is what can be called {\em CS-mod-residual.} It computes
\begin{eqnarray}
&& \hat{x}_T = \hat{\mu}_T,  \   \hat{x}_{T^c}=\hat{b}_c,  \ \text{where $\hat{b}_c$ solves}  \nn \\
&& \min_{b_c} \ \ \frac{1}{2}\|y-A_T\hat{\mu}_T-A_{T^c}b_c\|_2^2+\gamma \|b_c\|_1
\label{CSmodresidual}
\end{eqnarray}
where $b_c$ stands for $(b)_{T^c}$. This is solving a sparse recovery problem on $T^c$, i.e. it is implicitly assuming that $x_T$ is either equal to $\hat{\mu}_T$ or very close to it. Thus, this also works only when the signal value prior is very strong.

Both CS-residual and CS-mod-residual can be interpreted as extensions of BPDN, and \cite[Theorem 8]{justrelax} can be used to bound their error. In either case, the bound will contain terms proportional to $\|(x_T - \hat{\mu}_T)\|_2$ and as a result, it will be large whenever the prior is not strong enough\footnote{In either case, one can assume that $(x-\hat\mu)$ is supported on $\Delta$ and the ``noise" is $w+A_T(x_T-\hat{\mu}_T)$. Thus, CS-residual error can be bounded by $C(A,\Delta)(\|w\|_2 +\|A_T(x_T-\hat{\mu}_T)\|_2)$ while CS-mod-residual error can be bounded by $\|x_T-\hat{\mu}_T\|_2 + C(A_{T^c},\Delta)(\|w\|_2 +\|A_T(x_T-\hat{\mu}_T)\|_2)$.}. This is also seen from our simulation experiments shown in Fig. \ref{reconfig} where we provide comparisons for the case of good signal value prior (0.1\% error in initial signal estimate) and bad signal value prior (10\% error in initial signal estimate). We vary support errors from  5\% to 20\% misses, while keeping the extras fixed at 10\%.

Reg-mod-BPDN can also be confused with {\em modified-CS-residual} which computes\cite{modcsMRI}
\begin{eqnarray}
\hat{x} &=& \hat{\mu} + \hat{b}, \  \text{where $\hat{b}$ solves} \nn \\
&&  \min_{b} \ \ \frac{1}{2}\|y-A\hat{\mu}-A b\|_2^2+\gamma \|b_{T^c}\|_1
\label{modCSresidual}
\end{eqnarray}
This is indeed related to reg-mod-BPDN and in fact this inspired it. We studied this empirically in \cite{modcsMRI}. However, one cannot get good error bounds for it in any easy fashion. Notice that the minimization is over the entire vector $b$, while the $\ell_1$ cost is only on $b_{T^c}$. 


One may also consider solving the following variant of reg-mod-BPDN (we call this {\em reg-mod-BPDN-var}):
\begin{equation}\label{regmodBPDNvar}
\min_b \ \ \ \gamma\|b_{T^c}\|_1+\frac{1}{2}\|y-Ab\|_2^2+\frac{1}{2}\lambda\|b-\hat{\mu}\|_2^2
\end{equation}
Since $\muhat$ is supported on $T$, the regularization term can be rewritten as $\lambda\|b-\hat{\mu}\|_2^2  = \lambda\|b_T-\hat{\mu}_{T}\|_2^2+\lambda\|b_{T^c}\|_2^2$. Thus, in addition to the $\ell_1$ norm cost on $b_{T^c}$ imposed by the first term, this last term is also imposing an $\ell_2$ norm cost on it. If $\lambda$ is large enough, the $\ell_2$ norm cost will encourage the energy of the solution to be spread out on $T^c$, thus causing it to be less sparse. Since the true $x$ is very sparse on $T^c$ ($|\Delta|$ is small compared to the support size also), we will end up with a larger recovery error\footnote{In the limit if $\sqrt{\lambda/2}$ is much larger than $\gamma$, we may get a completely non-sparse solution.}.
[see Fig. \ref{reconfig}(a)].
However, if we compare the two approaches for compressible signal sequences, e.g. the larynx sequence, it is difficult to say which will be better [see Fig. \ref{larynxcompareregmodBPDN}].

Finally, one may solve the following ({\em we can call it reg-BPDN})
\begin{equation}\label{regmodBPDNvar_0}
\min_b \ \ \ \gamma\|b\|_1+\frac{1}{2}\|y-Ab\|_2^2+\frac{1}{2}\lambda\|b-\hat{\mu}\|_2^2
\end{equation}
This has two limitations. (1) Like CS-residual, this also does not use the fact that when $T$ is an accurate estimate of the true support, $(x)_{T^c}$ is much more sparse compared with the full $(x-\hat{\mu})$. (2) Its last term is the same as that of reg-mod-BPDN-var which also causes the same problem as above.%

\subsection{Paper Organization} 
We introduce reg-mod-BPDN in Sec. II. We obtain computable bounds on its reconstruction error in Sec. III. The simultaneous comparison of upper bounds of multiple approaches becomes difficult because their results hold under different sufficient
conditions. In Sec. IV, we address this issue by showing how to obtain a tighter error bound that also holds without any
sufficient conditions and is still computable. In both sections, the bounds for mod-BPDN and BPDN follow as direct corollaries.
%
In Sec V, the above result is used for easy numerical comparisons between the upper bounds of various
approaches -- reg-mod-BPDN, mod-BPDN, BPDN and LS-CS and for numerically evaluating tightness of the bounds with both Gaussian measurements and partial Fourier measurements.
We also provide reconstruction error comparisons with CS-residual, LS-CS, KF-CS, CS-mod-residual, mod-CS-residual and reg-mod-BPDN-var, as well as with weighted $\ell_1$, mod-BPDN and BPDN for (a) static sparse recovery from random-Gaussian measurements; and for (b) recovering a larynx image sequence from simulated MRI measurements.
Conclusions are given in Sec. VI.


\section{Regularized Modified-BPDN (Reg-mod-BPDN)}
Consider the sparse recovery problem when partial support knowledge is available. As explained earlier, one can use mod-BPDN given in (\ref{mb}). When the support estimate is accurate, i.e. $|\Delta|$ and $|\Delta_e|$ are small, mod-BPDN provides accurate recovery with fewer measurements than what BPDN needs. However, it puts no cost on $b_T$ except the cost imposed by the data term. Thus, when very few measurements are available or when the noise is large, $b_T$ can become larger than required (in order
to reduce the data term). A similar, though lesser, bias will occur with weighted $\ell_1$ also when $\gamma' < \gamma$.
To address this, when reliable prior signal value knowledge is available, we can instead solve
\begin{equation}\label{regmodBPDN}
\min_b \ \ \ L(b)\triangleq \gamma\|b_{T^c}\|_1+\frac{1}{2}\|y-Ab\|_2^2+\frac{1}{2}\lambda\|b_T-\hat{\mu}_T\|_2^2
\end{equation}
which we call {\em reg-mod-BPDN}. Its solution, denoted by $\hat{x}$, serves as the reconstruction of the unknown signal, $x$. Notice that the first term helps to find the solution that is sparsest outside $T$, the second term imposes the data constraint while the third term imposes closeness to $\hat\mu$ along $T$.

Mod-BPDN is the special case of (\ref{regmodBPDN}) when $\lambda = 0$.  BPDN is also a special case with $\lambda = 0$ and $T = \emptyset$  (so that $\Delta = N$).

\subsection{Limitations and Assumptions}
A limitation of adding the regularizing term, $\lambda\|b_T-\hat{\mu}_T\|_2^2$ is as follows.
It encourages the solution to be close to $(\muhat)_{\Delta_e}$ which is not zero. As a result, $(\xhat)_{\Delta_e}$ will also not be zero (except if $\lambda$ is very small) even though $(x)_{\Delta_e}=0$. Thus,  even in the noise-free case, reg-mod-BPDN will not achieve exact reconstruction. In both noise-free and noisy cases, if $(\muhat)_{\Delta_e}$ is large,  $(\xhat)_{\Delta_e}$ being close to $(\muhat)_{\Delta_e}$ can result in large error. Thus, we need the assumption that $(\muhat)_{\Delta_e}$ is small.


For the reason above, when we estimate the support of $\xhat$, we need to use a nonzero threshold, i.e. compute
\bea
\hat{N} = \{i: |\xhat_i| > \rho \}
\eea
with a $\rho > 0$. {We note that thresholding as above is done {\em only} for support estimation and not for improving the actual reconstruction. Support estimation is required in dynamic reg-mod-BPDN (described below) where we use the support estimate from the previous time instant as the support knowledge, $T$, for the current time.}



In summary, to get a small error reconstruction, reg-mod-BPDN requires the following (this can also be seen from the result of Theorem 1):%
\ben
\item $T$ is a good estimate of the true signal's support, $N$, i.e. $|\Delta|$ and $|\Delta_e|$ are small compared to $|N|$; and
\item $\muhat_T$ is a good estimate of $x_T$. For $i \in \Delta_e$, this implies that  $|\muhat_i|$ is close to zero (since $x_i = 0$ for $i \in \Delta_e$).
\item 
    For accurate support estimation, we also need that most nonzero elements of $x$ are larger than $\max_{i \in \Delta_e} |\muhat_i|$ (for exact support estimation, we need this to hold for all nonzero elements of $x$).
\een
The smallest nonzero elements of $x$ are usually on the set $\Delta$. In this case, the third assumption is equivalent to requiring that most elements of $x_\Delta$ are larger than  $\max_{i \in \Delta_e} |\muhat_i|$.

\subsection{Dynamic Reg-Mod-BPDN for Recursive Recovery} \label{dynreg}

An important application of reg-mod-BPDN is for recursively reconstructing a time sequence of sparse signals from undersampled measurements, e.g. for dynamic MRI. To do this, at time $t$ we solve (\ref{regmodBPDN}) with $T=\Nhat_{t-1}$, $(\muhat)_T = (\xhat_{t-1})_T$ and $(\muhat)_{T^c} = \mathbf{0}$. Here $\Nhat_{t-1}$ is the support estimate of the previous reconstruction, $\xhat_{t-1}$.
At the initial time, $t=0$, we can either initialize with BPDN, or with mod-BPDN using $T$ from prior knowledge, e.g. for wavelet sparse images, $T$ could be the set of indices of the approximation coefficients. We summarize the stepwise dynamic reg-mod-BPDN approach in Algorithm \ref{recursivealgo}. Notice that at $t=0$, one may need more measurements since the prior knowledge of $T$ may not be very accurate. Hence, we use $y_0 = A_0 x_0+w_0$ where $A_0$ is an $n_0 \times m$ measurement matrix with $n_0 > n$.

In Algorithm \ref{recursivealgo}, we should reiterate that for support estimation, we need to use a threshold $\rho > 0$. The threshold should be large enough so that most elements of $\Delta_{e,t}:=T \setminus N_{t} = \Nhat_{t-1} \setminus N_t$ do not get detected into the support.


We briefly discuss here the stability of dynamic reg-mod-BPDN (reconstruction error and support estimation errors bounded by a time-invariant and small value at all times). Using an approach similar to that of \cite{regmodBPDNstability}, it should be possible to show the following.
%
If (i) $\rho$ is large enough (so that $\Nhat_t$ does not falsely detect any element that got removed from $N_t$);
(ii) the newly added elements to the current support, $N_t$, either get added at a large enough value to get detected immediately,
or within a finite delay their magnitude becomes large enough to get detected;  and (iii) the matrix $A$ satisfies certain  conditions (for a given support size and support change size);  reg-mod-BPDN will be stable.

\begin{algorithm}[h!]
\caption{{\bf \small Dynamic Reg-mod-BPDN}}
At $t=0$, compute $\hat{x}_0$ as the solution of $\min_{b} \ \ \ \gamma\|(b)_{T^c}\|_1+\frac{1}{2}\|y_0-Ab\|_2^2$,
where $T$ is either empty or is available from prior knowledge.
Compute $\hat{N}_0 = \{i \in [1,...,m] : |(\hat{x}_{0})_i| > \rho \}$. Set $T \leftarrow \hat{N}_0$ and
 $(\muhat)_T \leftarrow (\hat{x}_{0})_T$
%
For $t>0$, do
\ben

\item {\em Reg-Mod-BPDN. } Let $T = \hat{N}_{t-1}$ and let  $\hat\mu_T = (\hat{x}_{t-1})_T$. 
Compute $\hat{x}_{t}$ as the solution of  (\ref{regmodBPDN}).
\label{step1recursive}

\item {\em Estimate Support. } $\hat{N}_t=\{i \in [1,...,m] : |(\hat{x}_{t})|_i > \rho \}$.

\item Output the reconstruction $\hat{x}_{t}$.
\een
Feedback $\Nhat_t$ and $\hat{x}_t$; increment $t$, and go to step \ref{step1recursive}.
\label{recursivealgo}
\end{algorithm}

\section{Bounding the Reconstruction Error}
In this section, we bound the reconstruction error of reg-mod-BPDN. Since mod-BPDN and BPDN are special cases, their results follow as direct corollaries. The result for BPDN is the same as \cite[Theorem 8]{justrelax}. In Sec. III-A, we define the terms needed to state our result. In III-B we state our result and discuss its implications. In III-C, we give the proof outline.

\subsection{Definitions}

We begin by defining the function that we want to minimize as
\begin{equation}
L(b) \triangleq  L_1(b)+\gamma\|b_{T^c}\|_1 \label{regmodBPDNfunction}
\end{equation}
where
\begin{equation}
L_1(b) \triangleq \frac{1}{2}\|y-Ab\|_2^2+\frac{1}{2}\lambda\|b_T-\hat{\mu}_T\|_2^2
\end{equation}
contains the two $\ell_2$ norm terms (data fidelity term and the regularization term). If we constrain $b$ to be supported on $T \cup S$ for some $S \subset T^c$, then the minimizer of $L_1(b)$ will be the regularized least squares (LS) estimator obtained when we put a weight $\lambda$ on $\|b_T-\hat{\mu}_T\|_2^2$ and a weight zero on $\|b_S-\hat{\mu}_S\|_2^2$.

Let $S$ be a given subset of $\Delta$. Next, we define three matrices which will be frequently used in our results. Let
\begin{eqnarray}
Q_{T,\lambda}(S) & \triangleq & {A_{T\cup S}}'A_{T\cup S}+\lambda  \left[
                                                 \begin{array}{cc}
                                                   I_{T}\ & \mathbf{0}_{T,S} \\
                                                   \mathbf{0}_{S,T}\ & \mathbf{0}_{S,S} \\
                                                 \end{array}
                                               \right] \label{def_Q} \\
M_{T,\lambda} & \triangleq & I-A_T({A_T}'A_T+\lambda I_{T})^{-1}{A_T}'  \\
P_{T,\lambda}(S)& \triangleq & ({A_{S}}'M_{T,\lambda}A_{S})^{-1} \label{def_P}
\end{eqnarray}
where $I_T$ is a $|T|\times |T|$ identity matrix and $\mathbf{0}_{T,S}$, $\mathbf{0}_{S,T}$, $\mathbf{0}_{S,S}$ are
all zeros matrices with sizes $|T|\times |S|$, $|S|\times |T|$ and $|S|\times |S|$.


\begin{assum}
We assume that  $Q_{T,\lambda}(\Delta)$ is invertible. This implies that, for any $S \subseteq \Delta$, the functions $L(b)$ and $L_1(b)$ are strictly convex over the set of all vectors supported on $T \cup S$.
\label{assum1}
\end{assum}
\begin{proposition}
When $\lambda > 0$, $Q_{T,\lambda}(S)$ is invertible if $A_S$ has full rank. When $\lambda =0$ (mod-BPDN), this will hold if $A_{T \cup S}$ has full rank.
\end{proposition}
The proof is easy and is given in Appendix \ref{Appprop1}.

Let $S \subseteq \Delta$. Consider minimizing $L(b)$ over $b$ supported on $T \cup S$. When $b_{(T \cup S)^c} = 0$ and Assumption \ref{assum1} holds, $L(b_{T \cup S})$ is strictly convex and thus has a unique minimizer. The same holds for $L_1(b_{T \cup S})$. Define their respective unique minimizers as%
\begin{eqnarray}
d_{T,\lambda}(S) &\triangleq& \arg \min_b \ L(b)  \ \ \text{ subject to }   \ \ b_{(T\cup S)^c}=\mathbf{0} \ \ \label{regminimizer} \\
c_{T,\lambda}(S) &\triangleq& \arg \min_b \ L_1(b)  \ \ \text{ subject to }   \ \ b_{(T\cup S)^c}=\mathbf{0} \ \
\end{eqnarray}
As explained earlier, $c_{T,\lambda}(S)$ is the regularized LS estimate of $x$ when assuming that $x$ is supported on $T \cup S$ and with the weights mentioned earlier. It is easy to see that
\begin{eqnarray}\label{reglssolution}
\ [c_{T,\lambda}(S)]_{T\cup S} &=&Q_{T,\lambda}(S)^{-1} \left({A_{T\cup S}}'y + \left[
                                                                                        \begin{array}{c}
                                                                                         \lambda \hat{\mu}_{T} \\
                                                                                          \mathbf{0}_{S} \\
                                                                                        \end{array}
                                                                                      \right] \right) \nonumber \\
\ [c_{T,\lambda}(S)]_{(T\cup S)^c} &=& \mathbf{0} \label{def_c}
\end{eqnarray}

In a fashion similar to \cite{justrelax}, define
\begin{eqnarray}
ERC_{T,\lambda}(S)& \triangleq & 1-\max_{\omega \notin
T\cup S}\|P_{T,\lambda}(S){A_{S}}'M_{T,\lambda}A_{\omega}\|_1  
\end{eqnarray}
This is different from the ERC of \cite{justrelax} but simplifies to it when $T=\emptyset$, $S=N$ and $\lambda=0$. In \cite{justrelax}, the ERC,
which in our notation is $ERC_{\emptyset,0}(N)$, being strictly positive, along with $\gamma$ approaching zero, ensured exact recovery of BPDN
in the noise-free case. Hence, in \cite{justrelax}, ERC was an acronym for {\em Exact Recovery Coefficient}. In this work, the same holds for
mod-BPDN. If $ERC_{T,0}(\Delta)>0$, the solution of mod-BPDN approaches the true $x$ as $\gamma$ approaches zero.
We explain this further in Remark \ref{erc_modcs} below. However, no similar claim can be made for reg-mod-BPDN. On the other hand,
for the reconstruction error bounds, ERC serves the exact same purpose for reg-mod-BPDN as it does for BPDN in \cite{justrelax}:
$ERC_{T,\lambda}(\Delta) > 0$ and $\gamma$ greater than a certain lower bound ensures that the reg-mod-BPDN (or mod-BPDN) error can be bounded by modifying the approach of \cite{justrelax}.

\subsection{Reconstruction error bound}
The reconstruction error can be bounded as follows.
\begin{theorem}
If $Q_{T,\lambda}(\Delta)$ is invertible, $ERC_{T,\lambda}(\Delta) > 0$ and
\begin{equation}
\gamma \ge \gamma^*_{T,\lambda}(\Delta) \triangleq \frac{\|{A_{(T\cup \Delta)^c}}'(y-Ac_{T,\lambda}(\Delta))\|_{\infty}}{ERC_{T,\lambda}(\Delta)} \label{regmodBPDNcond}
\end{equation}
then,
\begin{enumerate}
\item $L(b)$ has a unique minimizer, $\hat{x}$.
\item The minimizer, $\hat{x}$, is equal to $d_{T,\lambda}(\Delta)$, and thus is supported on $T\cup \Delta$.
\item Its error can be bounded as
\begin{eqnarray}
& &   \|x-\hat{x}\|_2 \le \gamma \sqrt{|\Delta|} f_1(\Delta)+\lambda f_2(\Delta)\|x_T-\hat\mu_T\|_2\nn \\
& &  \hspace{18mm}   +f_3(\Delta)\|w\|_2 \nonumber\\
& &\text{where     }  \nn \\
& & f_1(\Delta) \triangleq  \nn \\
 & &  \sqrt{\|({A_T}'A_T + \lambda I_{T})^{-1}   {A_T}'A_{\Delta}P_{T,\lambda}(\Delta)\|_{2}^2 +  \|P_{T,\lambda}(\Delta)\|_{2}^2}, \nonumber\\
& & f_2(\Delta) \triangleq   \|Q_{T,\lambda}(\Delta)^{-1}\|_2,    \nonumber\\
& &  f_3(\Delta) \triangleq \|Q_{T,\lambda}(\Delta)^{-1}{A_{T\cup \Delta}}'\|_2,  \ \hspace{-8mm} \label{loosebound}
\end{eqnarray}
$P_{T,\lambda}(\Delta)$ is defined in (\ref{def_P}) and $Q_{T,\lambda}(\Delta)$ in (\ref{def_Q}).
\end{enumerate}
\label{thm1}
\end{theorem}

\begin{corollary}[corollaries for mod-BPDN and BPDN]
The result for mod-BPDN follows by setting $\lambda=0$ in Theorem \ref{thm1}. The result for BPDN follows by setting $\lambda=0$, $T=\emptyset$ (and so $\Delta=N$). This result is the same as \cite[Theorem 8]{justrelax}.
\end{corollary}

\begin{remark}[smallest $\gamma$]
Notice that the error bound above is an increasing function of $\gamma$. Thus $\gamma=\gamma^*_{T,\lambda}(\Delta)$ gives the smallest bound.%
\end{remark}

In words, Theorem 1 says that, if $Q_{T,\lambda}(\Delta)$ is invertible, $ERC_{T,\lambda}(\Delta)$ is positive, and $\gamma$ is large
enough (larger than $\gamma^*$), then  $L(b)$ has a unique minimizer, $\xhat$, and $\xhat$ is supported on $T \cup \Delta = N \cup \Delta_e$.
This means that the only wrong elements that can possibly be part of the support of $\xhat$ are elements of $\Delta_e$.
Moreover, the error between $\xhat$ and the true $x$ is bounded by a value that is small as long as the noise, $\|w\|_2$, is small,
the prior term, $\|x_T- \muhat_T\|_2$, is small and $\gamma^*_{T,\lambda}(\Delta)$ is small. By rewriting $y - Ac_{T,\lambda}(\Delta) = A(x - c_{T,\lambda}(\Delta)) + w$  and using Lemma 2 (given in the Appendix) one can upper bound $\gamma^*$ by terms that are increasing functions of $\|w\|_2$ and  $\|x_T- \muhat_T\|_2$. Thus, as long as these are small, the bound is small.

As shown in Proposition 1, $Q_{T,\lambda}(\Delta)$ is invertible if $\lambda>0$ and $A_\Delta$ is full rank or if $A_{T \cup \Delta}$ is full rank.


Next, we use the idea of \cite[Corollary 10]{justrelax} to show that $ERC_{T,0}(\Delta)$ is an {\em Exact Recovery Coefficient} for mod-BPDN.%
%
\begin{remark}[ERC and exact recovery of mod-BPDN]
For mod-BPDN, $c_{T,0}(\Delta)$ is the LS estimate when $x$ is supported on $T \cup \Delta$. Using (\ref{def_c}), (\ref{obsmod}), and the fact that $x$ is supported on $N \subseteq T \cup \Delta$, it is easy to see that in the noise-free  ($w=0$) case, $c_{T,0}(\Delta) = x_{T \cup \Delta}$. Hence the numerator of $\gamma^*_{T,0}(\Delta)$ will be zero. Thus, using Theorem 1, if $ERC_{T,0}(\Delta)>0$, the mod-BPDN error satisfies $\|x- \xhat\|_2 \le \gamma \sqrt{|\Delta|} f_1(\Delta)$. Thus the mod-BPDN solution, $\xhat$, will approach the true $x$ as $\gamma$ approaches zero.
Moreover, as long as $\gamma < \frac{\min_{i \in N}|x_i|}{\sqrt{|\Delta|} f_1(\Delta)}$, at least the support of $\xhat$ will equal the true support, $N$ \footnote{If we bounded the $\ell_\infty$ norm of the error as done in \cite{justrelax} we would get a looser upper bound on the allowed $\gamma$'s for this.}.
\label{erc_modcs}
\end{remark} 


We show a numerical comparison of the results of reg-mod-BPDN, mod-BPDN and BPDN in Table I (simulation details given in Sec. V).
Notice that BPDN needs $90\%$ of the measurements for its sufficient conditions to start holding (ERC to become positive) whereas mod-BPDN only needs $19\%$.
Moreover, even with $90\%$ of the measurements, the ERC of BPDN is just positive and very small. As a result, its error bound is
large ($27\%$ normalized mean squared error (NMSE)). Similarly, notice that mod-BPDN needs $n \ge 19\% m$ for its sufficient conditions to start holding ($A_{T \cup \Delta}$ to become full rank which is needed for $Q_{T,0}(\Delta)$ to be invertible). For reg-mod-BPDN which only needs $A_\Delta$ to be full rank, $n=13\% m$ suffices.

\begin{remark}
A sufficient conditions comparison only provides a comparison of when a given result can be applied to provide a bound on the reconstruction error. In other words, it tells us under what conditions we can guarantee that the reconstruction error of a given approach will be small (below a bound). Of course this does not mean that we cannot get small error even when the sufficient condition does not hold, e.g., in simulations, BPDN provides a good reconstruction using much less than 90\% of the measurements. However, when $n<90\% m$ we cannot bound its reconstruction error using Theorem 1 above.
\end{remark}



\subsection{Proof Outline}
To prove Theorem \ref{thm1}, we use the following approach motivated by that of \cite{justrelax}.
\begin{enumerate}
\item  We first bound $\|d_{T,\lambda}(\Delta)-c_{T,\lambda}(\Delta)\|_2$ by simplifying the necessary and sufficient condition for it to be the minimizer of $L(b)$ when $b$ is supported on $T \cup \Delta$.
    This is done in Lemma 1 in Appendix \ref{App3lemmas}.

\item  We bound $\|c_{T,\lambda}(\Delta)-x\|_2$ using the expression for $c_{T,\lambda}(\Delta)$ in (\ref{reglssolution}) and substituting $y = A_{T \cup \Delta} x_{T \cup \Delta} + w$ in it (recall that $x$ is zero outside $T \cup \Delta$). This is done in Lemma 2 in Appendix \ref{App3lemmas}.

\item We can bound $\|d_{T,\lambda}(\Delta)-x\|_2$ using the above two bounds and the triangle inequality.

\item We use an approach similar to \cite[Lemma 6]{justrelax} to find the sufficient conditions under which $d_{T,\lambda}(\Delta)$ is also the unconstrained unique minimizer of $L(b)$, i.e. $\hat{x} = d_{T,\lambda}(\Delta)$. This is done in Lemma 3 in Appendix \ref{App3lemmas}.
\end{enumerate}
The last step (Lemma 3) helps prove the first two parts of Theorem 1. Combining the above four steps, we get the third part (error bound).
We give the lemmas in Appendix \ref{App3lemmas}. They are proved in Appendix \ref{Applemma1proof}, \ref{Applemma2proof} and \ref{Applemma3proof}.

Two key differences in the above approach with respect to the result of \cite{justrelax} are
\begin{itemize}
 \item $c_{T,\lambda}(\Delta)$ is the regularized LS estimate instead of the LS estimate in \cite{justrelax}.
 This helps obtain a better and simpler error bound of reg-mod-BPDN than when using the LS estimate. Of course, when $\lambda=0$ (mod-BPDN or BPDN),
 $c_{T,0}(\Delta)$ is just the LS estimate again.

\item For reg-mod-BPDN (and also for mod-BPDN), the subgradient set of the $\ell_1$ term
 is $\partial \|b_{T^c}\|_1|_{b=d_{T,\lambda}(\Delta)}$ and so any $\phi$ in this set is zero on $T$,
 and only has $\|\phi_{\Delta}\|_{\infty}\le 1$. Since $|\Delta| \ll |N|$, this helps to get a tighter bound
 on $\|c_{T,\lambda}(\Delta) - d_{T,\lambda}(\Delta)\|_2$ in step 1 above as compared to that for
 BPDN \cite{justrelax} (see proof of Lemma 1 for details).

\end{itemize}


\section{Tighter Bounds without Sufficient Conditions}
The problem with the error bounds for reg-mod-BPDN, mod-BPDN, BPDN or LS-CS \cite{LSCSbound} is that they all hold under different sufficient conditions. This makes it difficult to compare them. Moreover, the bound is particularly loose when $n$ is such that the sufficient conditions just get satisfied. This is because the ERC is just positive but very small (resulting in a very large $\gamma^*$ and hence a very large bound).
To address this issue, in this section, we obtain a bound that holds without any sufficient conditions and that is also tighter, while still being computable. 
The key idea that we use is as follows:
 \begin{itemize}
 \item we modify Theorem 1 to hold for ``sparse-compressible" signals \cite{LSCSbound}, i.e. for sparse signals, $x$, in which some nonzero coefficients out of the set $\Delta$ are small (``compressible") compared to the rest; and then
 \item we minimize the resulting bound over all allowed split-ups of $x$ into non-compressible and compressible parts.%
 \end{itemize}

 Let $\tilde{\Delta} \subseteq \Delta$ be such that the conditions of Theorem 1 hold for it. Then the first step involves modifying Theorem 1 to bound the error for reconstructing $x$ when we treat $x_{\Delta \setminus \tilde{\Delta}}$ as the ``compressible" part. The main difference here is in bounding $\|c_{T,\lambda}(\tilde{\Delta})-x\|_2$ which now has a larger bound because of $x_{\Delta \setminus \tilde{\Delta}}$. We do this in Lemma \ref{lemma4} in the Appendix \ref{Applemma4}. Notice from the proofs of Lemma 1 and Lemma 3 in Appendix \ref{Applemma1proof} and \ref{Applemma3proof} that nothing in their result changes if we replace $\Delta$ by a $\tDelta \subseteq \Delta$. Combining Lemma 4 with Lemmas 1 and 3 applied for $\tDelta$ instead of $\Delta$ leads to the following corollary.
\begin{corollary}
Consider a $\tDelta \subseteq \Delta$. If $Q_{T,\lambda}(\tilde{\Delta})$ is invertible, $ERC_{T,\lambda}(\tilde{\Delta})>0$, and $\gamma=\gamma^*_{T,\lambda}(\tilde{\Delta})$, then
\begin{equation}
\|x-\hat{x}\|_2 \le f(T,\lambda,\Delta,\tilde{\Delta},\gamma^*_{T,\lambda}(\tilde{\Delta}))
\end{equation}
where
\begin{eqnarray}
f(T,\lambda,\Delta,\tilde{\Delta},\gamma) & \triangleq& \gamma \sqrt{|\tilde{\Delta}|} f_1(\tDelta)+ \lambda  f_2(\tDelta) \|x_{T}-\hat{\mu}_{T}\|_2 \nonumber\\ & &   + f_3(\tDelta)\|w\|_2  +
  f_4(\tDelta) \|x_{\Delta \setminus \tilde{\Delta}}\|_2,  \label{def_f} \\
f_4(\tDelta) & \triangleq & \sqrt{\|Q_{T,\lambda}(\tilde{\Delta})^{-1}{A_{T\cup \tilde{\Delta}}}'A_{\Delta \setminus \tilde{\Delta}}\|_2^2+1}, \ \ \ \ \
\label{def_f4}
\end{eqnarray}
$f_1(\cdot)$,$f_2(\cdot)$, $f_3(\cdot)$ are defined in (\ref{loosebound}) and $\gamma^*_{T,\lambda}(\tilde{\Delta})$ in (\ref{regmodBPDNcond}).
\label{cor2}
\end{corollary}

{\em Proof: } The proof is given in Appendix \ref{Appcor1proof}.

In order to get a bound that depends only on $\|x_T-\hat{\mu}_T\|_2$, $\|x_{\Delta \setminus \tDelta}\|_2$, the noise, $w$, and the sets $T,\Delta,\Delta_e$, we can further bound $\gamma^*_{T,\lambda}(\tilde{\Delta})$ by rewriting $y-Ac_{T,\lambda}(\tilde{\Delta}) = A(x-c_{T,\lambda}(\tilde{\Delta})) + w$ and then bounding $\|x-(c_{T,\lambda}(\tilde{\Delta}))\|_2$ using Lemma 4. Doing this gives the following corollary.
\begin{corollary}
If $Q_{T,\lambda}(\tilde{\Delta})$ is invertible, $ERC_{T,\lambda}(\tilde{\Delta})>0$, and $\gamma=\gamma^*_{T,\lambda}(\tilde{\Delta})$, then
\begin{equation}
\|x-\hat{x}\|_2 \le g(\tDelta) \label{gdefbound}
\end{equation}
where
\begin{eqnarray}
\hspace{-10mm} g(\tDelta) \sdefn g_1 \|x_T - \hat{\mu}_T\|_2 + g_2 \|w\|_2 + g_3 \|x_{\Delta \setminus \tDelta}\|_2+g_4 \hspace{7mm} \label{gbound} \\
g_1 \sdefn   \lambda f_2(\tDelta) (\frac{\sqrt{|\tDelta|}f_1(\tDelta) \text{maxcor}(\tDelta)}{ERC_{T,\lambda}(\tDelta)}+1 ), \nn \\
g_2 \sdefn  \frac{\sqrt{|\tDelta|}f_1(\tDelta)f_3(\tDelta) \text{maxcor}(\tDelta) }{ERC_{T,\lambda}(\tDelta)}+f_3(\tDelta), \nn \\
g_3 \sdefn \frac{\sqrt{|\tDelta|}f_1(\tDelta)f_4(\tDelta) \text{maxcor}(\tDelta) }{ERC_{T,\lambda}(\tDelta)}+f_4(\tDelta), \nn \\
g_4 \sdefn \frac{\sqrt{|\tDelta|}\|{A_{(T\cup \tDelta)^c}}'w\|_{\infty}f_1(\tDelta)}{ERC_{T,\lambda}(\tDelta)}, \quad \nn\\
\text{maxcor}(\tDelta)\sdefn \max_{i\notin (T\cup \tDelta)^c}\|{A_i}'A_{T\cup \Delta}\|_2, \nn
\end{eqnarray}
$f_1(\cdot)$,$f_2(\cdot)$, $f_3(\cdot)$ and $f_4(\cdot)$ are defined in (\ref{loosebound}) and (\ref{def_f4}), and $\gamma^*_{T,\lambda}(\tilde{\Delta})$ in (\ref{regmodBPDNcond}).
\label{corol2}
\end{corollary}

{\em Proof: } The proof is given in Appendix \ref{Appcor2proof}.

Using the above corollary and minimizing over all allowed $\tDelta$'s, we get the following result.
\begin{theorem}
Let
\begin{equation}
\tilde{\Delta}^* \triangleq \ \ \arg \min_{\hspace{-8mm}\tilde{\Delta}\in \mathcal{G}} g(\tDelta)
\end{equation}
where
\begin{equation}
\mathcal{G} \triangleq \{\tilde{\Delta}:\tilde{\Delta}\subseteq \Delta, ERC_{T,\lambda}(\tilde{\Delta})>0, Q_{T,\lambda}(\tilde{\Delta}) \text{ is invertible}\}
\end{equation}
If $\gamma=\gamma^*_{T,\lambda}(\tilde{\Delta}^*)$, then
\begin{enumerate}
\item $L(b)$ has a unique minimizer, $\hat{x}$, supported on $T\cup \tilde{\Delta}^*$.

\item The error bound is
\begin{equation}
\|x-\hat{x}\|_2\le g(\tDelta^*)
\label{regmodBPDNtightbound}
\end{equation}
\end{enumerate}
($\gamma^*_{T,\lambda}(\tilde{\Delta})$ is defined in (\ref{regmodBPDNcond})).
\label{thm2}
\end{theorem}

{\em Proof: } This result follows by minimizing over all allowed $\tDelta$'s from Corollary \ref{corol2}.

Compare Theorem 2 with Theorem 1. Theorem 1 holds only when the complete set $\Delta$ belongs to $\mathcal{G}$, whereas Theorem 2 holds always (we only need to set $\gamma$ appropriately). Moreover, even when $\Delta$ does belong to $\mathcal{G}$, Theorem 1 gives the error bound by choosing $\tilde{\Delta}^* = \Delta$. However, Theorem 2 minimizes over all allowed ${\tDelta}$'s, thus giving a tighter bound, especially for the case when the sufficient conditions of Theorem 1 just get satisfied and $ERC_{T,\lambda}(\Delta)$ is positive but very small. A similar comparison also holds for the mod-BPDN and BPDN results.

The problem with Theorem 2 is that its bound is not computable (the computational cost is exponential in $|\Delta|$). Notice that $g(\tDelta^*):=\min_{\tilde{\Delta}\in \mathcal{G}} g(\tDelta)$ can be rewritten as
\begin{eqnarray}
& & g(\tDelta^*)\triangleq \min_{\tilde{\Delta} \in \mathcal{G}} g(\tDelta) = \min_{0 \le k \le |\Delta|} \min_{\mathcal{G}_k} g(\tDelta) \ \ \text{where} \nn \\
& & \mathcal{G}_k \triangleq \mathcal{G}  \cap \{\tDelta \subseteq \Delta : |\tDelta| = k\}
\label{Gkdef}
\end{eqnarray}
Let $d:=|\Delta|$. The minimization over $\mathcal{G}_k$ is expensive since it requires searching over all ${d \choose k}$ size $k$ subsets of $\Delta$ to first find which ones belong to $\mathcal{G}_k$ and then find the minimum over all $\tDelta \subseteq \mathcal{G}_k$.  The total computation cost to do the former for all sets $\mathcal{G}_0, \mathcal{G}_1, \dots \mathcal{G}_d$ is $O(\sum_{k=0}^d {d \choose k}) = O(2^d)$, i.e. it is {\em exponential in $d$}. This makes the bound computation intractable for large problems.

\subsection{Obtaining a Computable Bound}
In most cases of practical interest, the term that has the maximum variability over different sets in $\mathcal{G}_k$ is $\|x_{\Delta \setminus \tDelta}\|_2$. The multipliers $g_1$, $g_2$, $g_3$ and $g_4$ vary very slightly for different sets in a given $\mathcal{G}_k$. Using this fact, we can obtain the following upper bound on $\min_{\mathcal{G}_k} g(\tDelta)$ which is only slightly looser and also holds without sufficient conditions, but is computable in polynomial time.

Define $\tDelta^{**}(k)$ and $B_k$ as follows
\begin{eqnarray}
\tDelta^{**}(k) \sdefn \arg \min_{ \{\tDelta \subseteq \Delta, |\tDelta|=k \} } \|x_{\Delta \setminus \tDelta}\|_2 \nn \\
B_k \sdefn \left\{ \begin{array}{cc}
                  g(\tDelta^{**}(k)) & \ \text{if} \ \tDelta^{**}(k) \in \mathcal{G}_k \\
                  \infty          & \ \text{otherwise} \
                  \end{array}
                  \right.
\label{defBk}
\end{eqnarray}
Then, clearly
\begin{eqnarray}
\min_{\mathcal{G}_k} g(\tDelta) \le B_k
\label{Bk_bnd}
\end{eqnarray}
since $\min_{\mathcal{G}_k} g(\tDelta) \le g(\tDelta)$ for any $\tDelta \in \mathcal{G}_k$ and it is also less than infinity.
For any $k$, the set $\tDelta^{**}(k)$ can be obtained by sorting the elements of $x_\Delta$ in decreasing order of magnitude and letting $\tDelta^{**}(k)$ contain the indices of the $k$ largest elements. Doing this takes $O(d \log d)$ time since sorting takes $O(d \log d)$ time. Computation of $B_k$ requires matrix multiplications and inversions which are $O(k^3)$. Thus, the total cost of doing this is at most $O(d^4)$ which is still polynomial in $d$.

 Therefore, we get the following bound that is {\em computable in polynomial time and that still holds without
 sufficient conditions and is much tighter than Theorem 1}.
\begin{theorem}
Let
\begin{eqnarray}
k_{\min} \sdefn \arg \min_{0 \le k \le |\Delta|} B_k  \quad \text{  and }\label{kmindef} \nn \\
\tDelta^{**} \sdefn \tDelta^{**}(k_{\min})
\end{eqnarray}
where $B_k$ and $\tDelta^{**}(k)$ are defined in (\ref{defBk}).
If $\gamma=\gamma^*_{T,\lambda}(\tDelta^{**})$,
\begin{enumerate}
\item $L(b)$ has a unique minimizer, $\hat{x}$, supported on $T\cup \tDelta^{**}$.
\item The error bound is
\begin{equation}
\|x-\hat{x}\|_2 \le  g(\tDelta^{**})
\label{regmodBPDNmediumtightbound}
\end{equation}
\end{enumerate}
($\gamma^*_{T,\lambda}(\tilde{\Delta})$ is defined in (\ref{regmodBPDNcond})).
\label{thm3}
\end{theorem}
\begin{corollary}[corollaries for mod-BPDN and BPDN]
The result for mod-BPDN follows by setting $\lambda=0$ in Theorem \ref{thm3}. The result for BPDN follows by setting $\lambda=0$, $T=\emptyset$ (and so $\Delta=N$) in Theorem \ref{thm3}.
\end{corollary}


When $n$ and $s \triangleq |N|$ are large enough, the above bound is either only slightly larger, or often actually equal, to that of Theorem 2 (e.g. in Fig. \ref{bound4comparefig}(a), $m=256$, $n =0.13m=33$, $s=0.1m=26$). The reason for the equality is that  the minimizing value of $k$ is the one that is small enough to ensure that $g_1, g_2, g_3,g_4$ are small. When $k$ is small, $g_1, g_2, g_3, g_4$, ${ERC}$ and $Q(\tDelta)$ have very similar values for all sets $\tDelta$ of the same size $k$. In (\ref{gbound}), the only term with significant variability for different sets $\tDelta$ of the same size $k$ is $\|x_{\Delta \setminus \tDelta}\|_2$. Thus,
(a) $\arg\min_{\mathcal{G}_k} g(\tDelta) = \arg\min_{\mathcal{G}_k} \|x_{\Delta \setminus \tDelta}\|_2$ and
(b) $\mathcal{G}_k$ is equal to $\{\tDelta \subseteq \Delta, |\tDelta|=k\}$.  
Thus, (\ref{Bk_bnd}) holds with equality and so the bounds from Theorems 3 and 2 are equal.
%
As $n$ and $s\triangleq |N|$ approach infinity, {\em it is possible to use a law of large numbers (LLN) argument to prove that both bounds will be equal with high probability (w.h.p.)}. The key idea will be the same as above: show that as $n,s$ go to infinity,  w.h.p., $g_1,g_2,g_3,g_4$, $Q$ and $ERC$ are equal for all sets $\tDelta$ of any given size $k$. We will develop this result in future work.


\section{Numerical Experiments}
In this section, we show both upper bound comparisons and actual reconstruction error comparisons. The upper bound comparison only tells us that the performance guarantees of reg-mod-BPDN are better than those for the other methods.  To actually demonstrate that reg-mod-BPDN outperforms the others, we need to compare the actual reconstruction errors. This section is organized as follows. After giving the simulation model in Sec V-A, we show the reconstruction error comparisons for recovering simulated sparse signals from random Gaussian measurements in Sec V-B. In Sec V-C, we show comparisons for recursive dynamic MRI reconstruction of a larynx image sequence. In this comparison, we also show the usefulness of the Theorem 3 in helping us select a good value of $\gamma$. In the last three subsections, we show numerical comparisons of the results of the various theorems. The upper bound comparisons of Theorem 3 and the comparison of the corresponding reconstruction errors suggests that the bounds for reg-mod-BPDN and BPDN are tight under the scenarios evaluated. Hence, they can be used as a proxy to decide which algorithm to use when. We show this for both random Gaussian and partial Fourier measurements.


\subsection{Simulation Model}
The notation $z = \pm a$  means that we generate each element of the vector $z$ independently and each is either $+a$ or $-a$ with probability 1/2. The notation $\nu \sim \n(0, \Sigma)$ means that $\nu$ is generated from a Gaussian distribution with mean 0 and covariance matrix $\Sigma$. We use $\lfloor a \rfloor$  to denote the largest integer less than or equal to $a$.  Independent and identically distributed is abbreviated as iid. Also, N-RMSE refers to the normalized root mean squared error.


We use the recursive reconstruction application \cite{LSCS,modcsjournal} to motivate the simulation model. In this case,
assuming that slow support and slow signal value change hold [see Fig. \ref{slowchange}], we can use the reconstructed value of the signal
at the previous time as $\muhat$ and its support as $T$. To simulate the effect of slow signal value change, we let $x_N = \mu_N + \nu$ where $\nu$
is a small iid Gaussian deviation and we let $\muhat_{T \cap N} =  \mu_{T \cap N}$ (and so $x_{T \cap N} =  \muhat_{T \cap N} + \nu_{T \cap N}$).

The extras set, $\Delta_e = T \setminus N$, contains elements that got removed from the support at the current time or at a few previous times (but so far did not get removed from the support estimate). In most practical applications, only small valued elements at the previous time get removed from the support and hence the magnitude of $\hat\mu$ on $\Delta_e$ will be small. We use $\beta_s$ to denote this small magnitude, i.e. we simulate $(\muhat)_{\Delta_e} = \pm \beta_s$.%

The misses' set at time $t$, $\Delta$, definitely includes the elements that just got added to the support at $t$ or the ones that previously got added but did not get detected into the support estimate so far. The new elements typically get added at a small value and their value slowly increases to a large one. Thus, elements in $\Delta$ will either have small magnitude (corresponding to the current newly added ones), or will have larger magnitude but still smaller than that of elements already in $N \cap T$.
To simulate this, we do the following. (a) We simulate the elements on $N \cap T$ to have large magnitude, $\beta_l$, i.e.
we let $(\mu)_{N \cap T} = \pm \beta_l$.
(b) We split the set $\Delta$ into two disjoint parts, $\Delta_1$ and $\Delta_2 = \Delta \setminus \Delta_1$. The set $\Delta_1$ contains the small (e.g. newly added) elements, i.e. $(\mu)_{\Delta_1} = \pm \beta_s$.
The set $\Delta_2$ contains the larger elements, though still with magnitudes smaller than those in $N \cap T$, i.e. $(\mu)_{\Delta_2} = \pm \beta_m$, where $\beta_l \ge \beta_m \ge \beta_s$.

In summary, we use the following simulation model.
\begin{eqnarray}
(x)_N \se (\mu)_N + \nu, \ \ \nu \sim \n(0,\sigma_p^2 I)  \nn \\
(x)_{N^c} \se 0  \label{xgen} \\
%
\text{where} \ \ (\mu)_{N \cap T} \se \pm \beta_l \nn \\
(\mu)_{\Delta_1} \se \pm \beta_s, \ \ (\mu)_{\Delta_2} = \pm \beta_m \nn \\
(\mu)_{N^c} \se 0 \label{mugen}
\end{eqnarray}
and
\begin{eqnarray}
(\hat\mu)_{T \cap N} \se (\mu)_{T \cap N}  = \pm \beta_l \nn \\
(\hat\mu)_{\Delta_e} \se  \pm \beta_s \nn \\
(\hat\mu)_{T^c} \se 0 \label{muhatgen}
\end{eqnarray}
%
We generate the support of $x$, $N$, of size $|N|$, uniformly at random from $[1,...,m]$. We generate $\Delta$ with size $|\Delta|$ and $\Delta_e$ with size $|\Delta_e|$ uniformly at random from $N$ and from $N^c$ respectively. The set $\Delta_1$ of size $|\Delta_1|=\lfloor |\Delta|/2\rfloor$ is generated uniformly at random from $\Delta$. The set $\Delta_2 = \Delta \setminus \Delta_1$. We let $T =  N \cup \Delta_e \setminus \Delta$. We generate $\mu$ and then $x$ using (\ref{mugen}) and (\ref{xgen}). We generate $\hat\mu$ using (\ref{muhatgen}).

In some simulations, we simulated the more difficult case where $\beta_m = \beta_s$. In this case, all elements on $\Delta$ were identically generated and hence we did not need $\Delta_1$.

\begin{figure*}[t!]
\subfigure[\small{$n=0.13m$,$\sigma_p^2=10^{-3}$,$\sigma_w^2=10^{-5}$}]{
\label{fig3a}
\includegraphics [width=9cm,height=6cm]{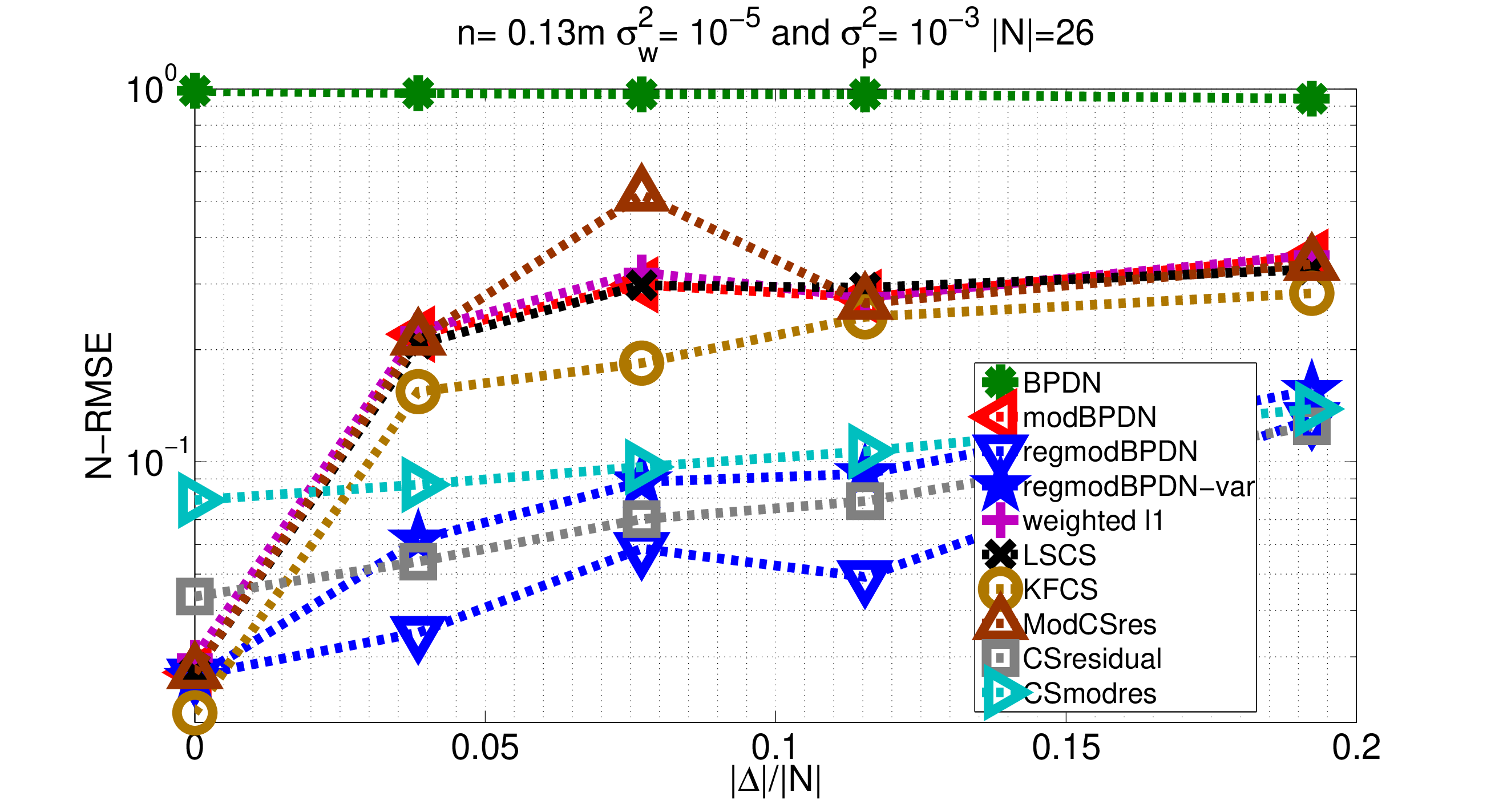}
}
\hspace{-5mm}
\subfigure[\small{$n=0.13m$,$\sigma_p^2=10^{-1}$,$\sigma_w^2=10^{-5}$}]{
\includegraphics [width=9cm,height=6cm]{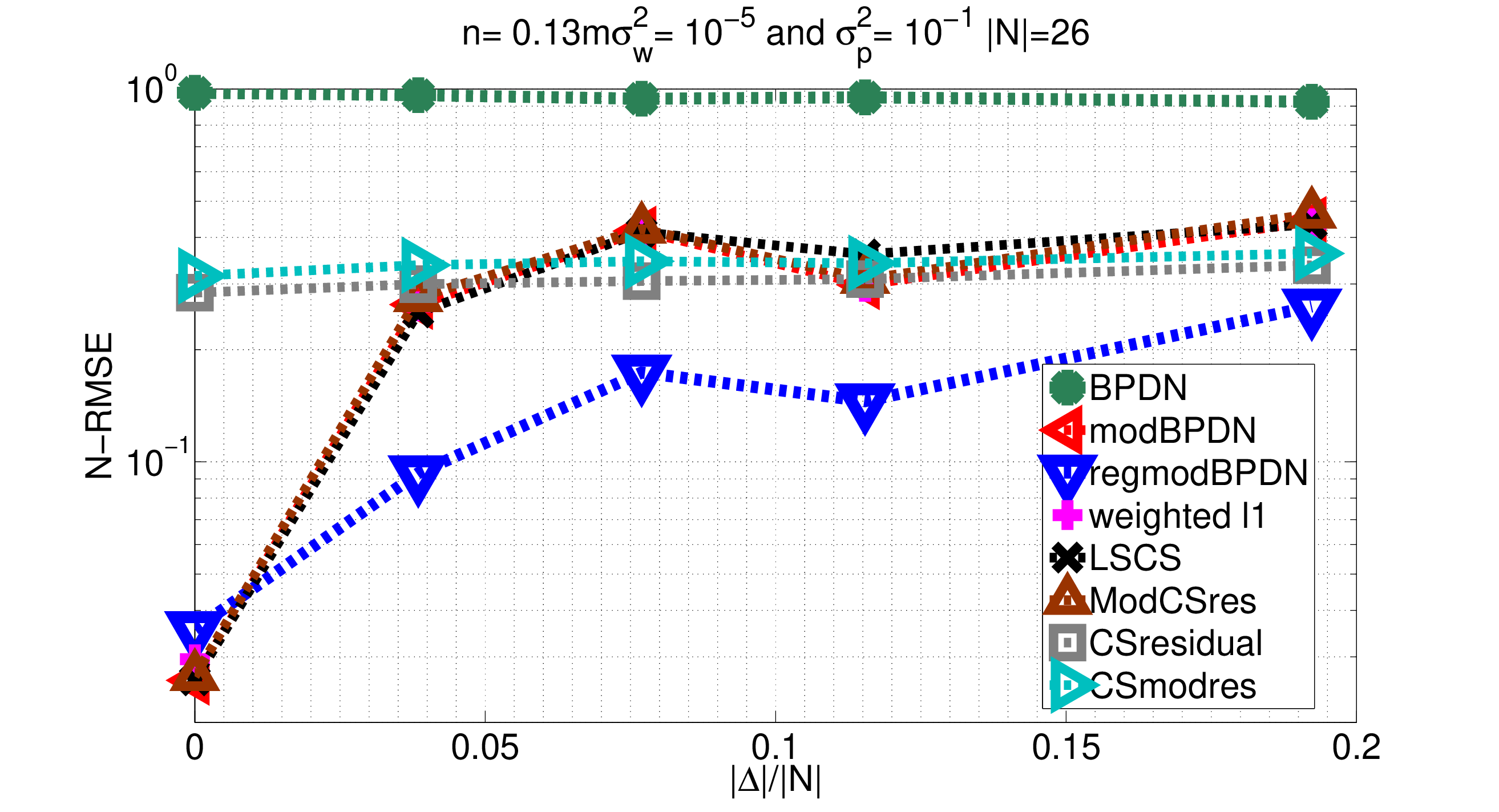}
}\\
\subfigure[\small{$n=0.3m$,$\sigma_p^2=10^{-3}$,$\sigma_w^2=10^{-4}$}]{
\includegraphics [width=9cm,height=6cm]{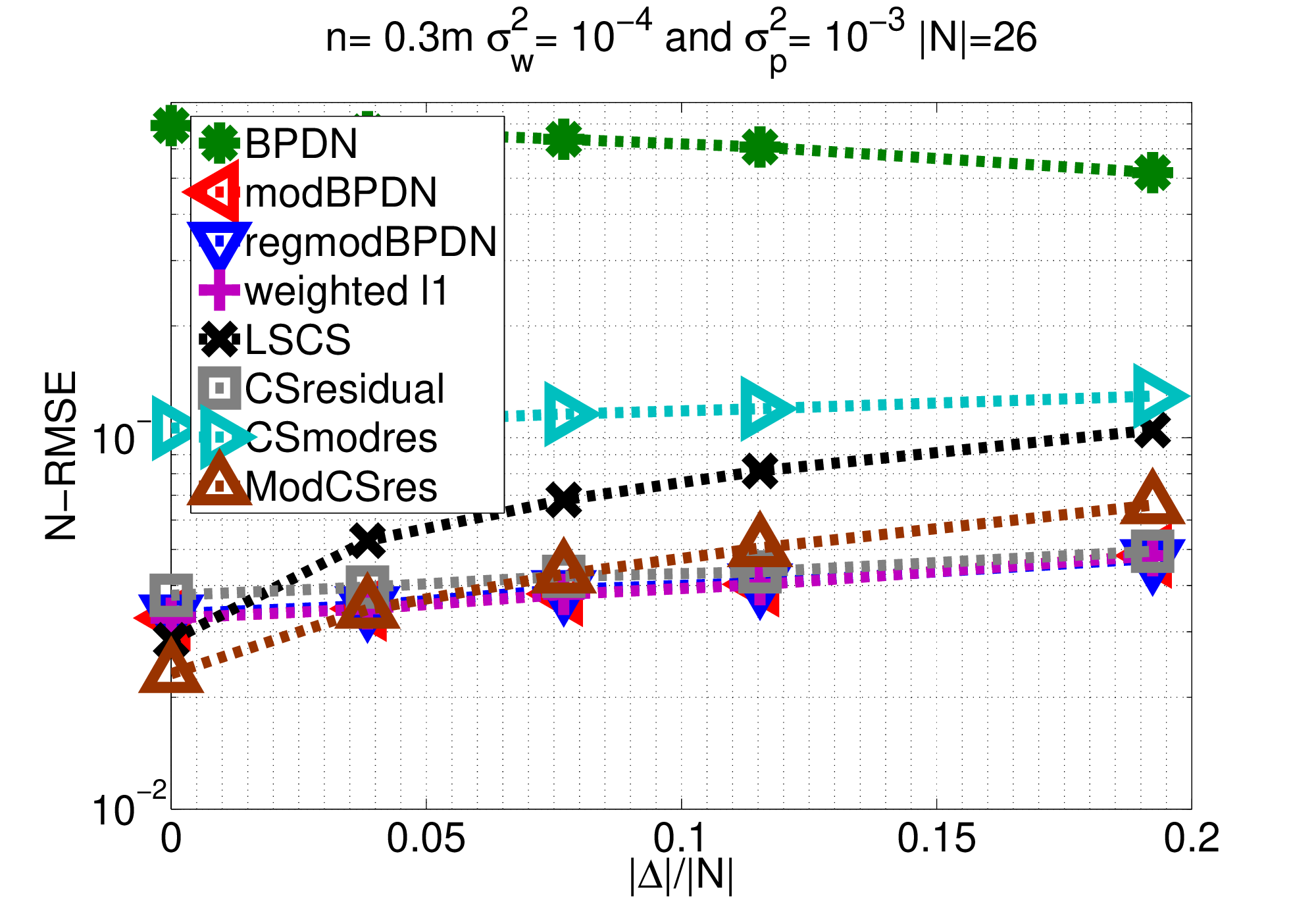}
}
\hspace{-5mm}
\subfigure[\small{$n=0.3m$,$\sigma_p^2=10^{-1}$,$\sigma_w^2=10^{-5}$}]{
\includegraphics [width=9cm,height=6cm]{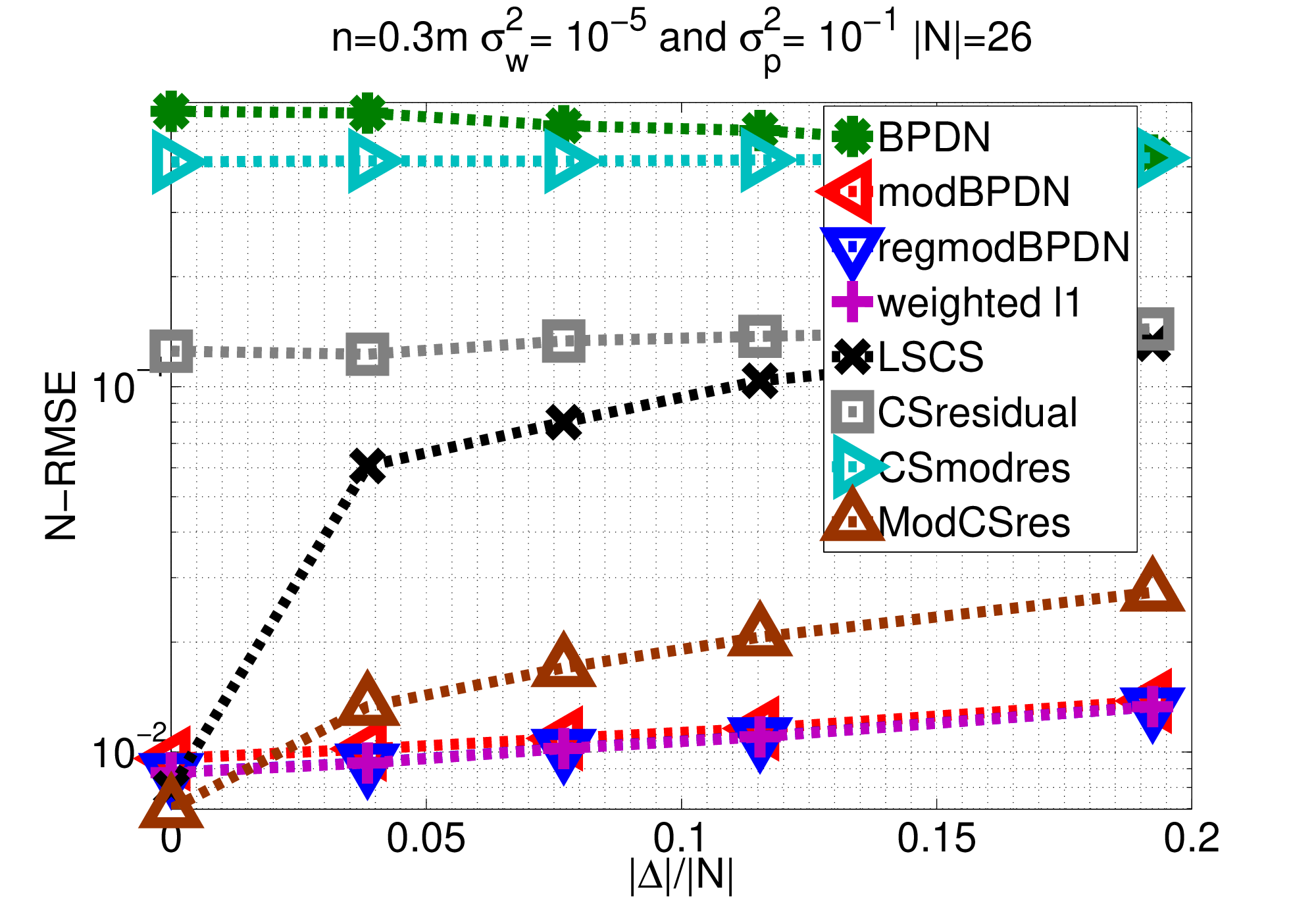}
}
\vspace{-1mm}
\caption{\small{The N-RMSE for reg-mod-BPDN, mod-BPDN, BPDN, LS-CS, KF-CS, weighted $\ell_1$, CS-residual, CS-mod-residual and modified-CS-residual are plotted.
For $n=0.13m$ , reg-mod-BPDN has smaller errors than those of mod-BPDN and the gap is larger when the signal estimate is good. For $n=0.3m$, the errors of reg-mod-BPDN, mod-BPDN and weighted $\ell_1$ are close and all small. }
} \label{reconfig}
\end{figure*}

\begin{figure}
\includegraphics [width=9cm,height=6cm]{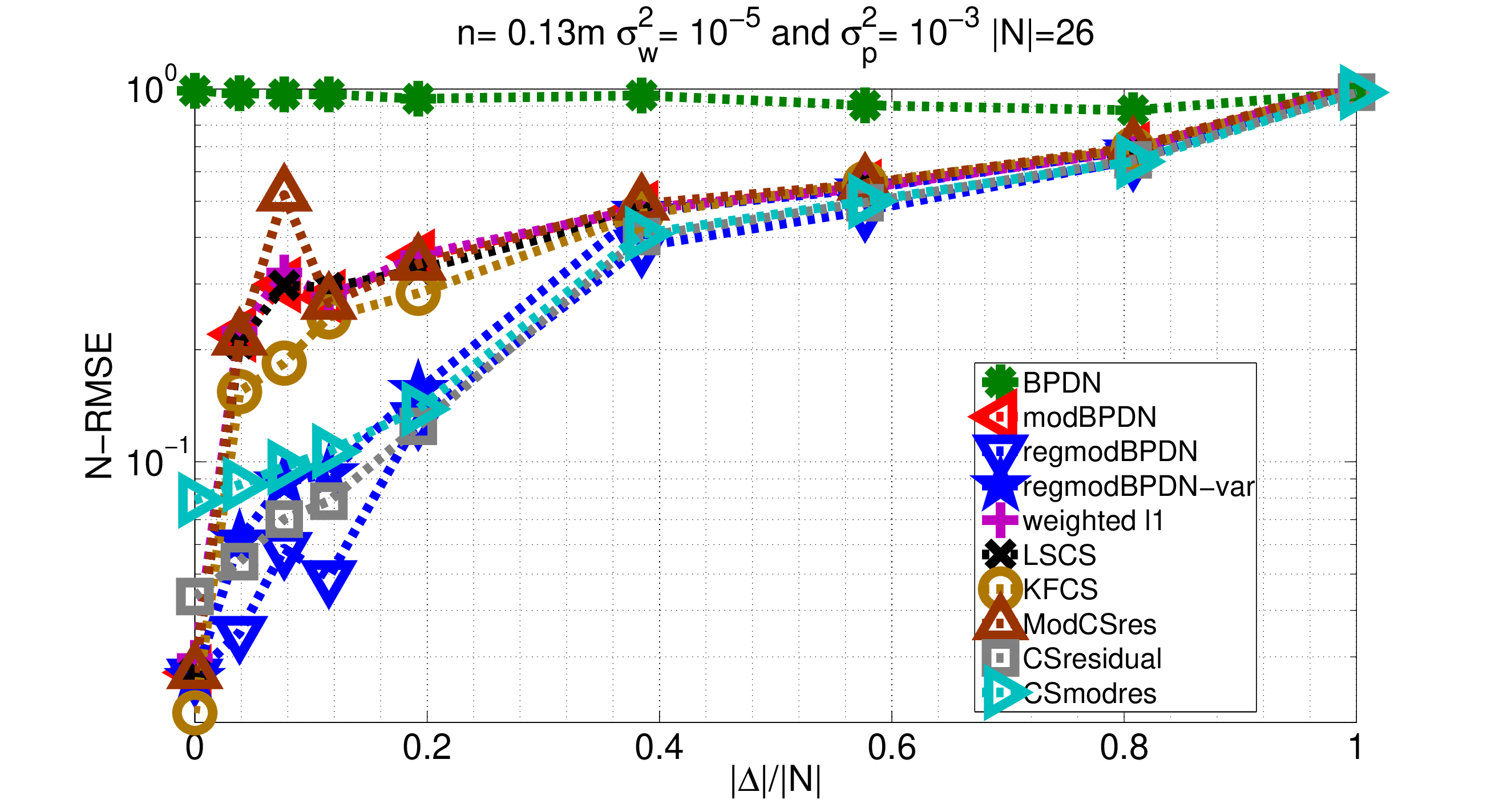}
\vspace{-1mm}
\caption{\small{{Plot of Fig \ref{fig3a} extended all the way to $|\Delta|/|N| = 1$ (which is the same as $\Delta = N$).  Notice that if $|\Delta_e|=0$, then the point $|\Delta|/|N|=1$ of reg-mod-BPDN (or of mod-BPDN) is the same as BPDN. But in our plot, $|\Delta_e|=3$ and hence the two points are different, even though the errors are quite similar.}
}}
\label{fig_extend}
\end{figure}

\subsection{Reconstruction Error Comparisons}
In Fig. \ref{reconfig}, we compare the Monte Carlo average of the reconstruction error of reg-mod-BPDN with that of mod-BPDN, BPDN, weighted $\ell_1$ \cite{weightedl1} given in (\ref{weightedl1}),
CS-residual given in (\ref{CSresidual}), CS-mod-residual given in (\ref{CSmodresidual}) and modified-CS-residual\cite{modcsMRI} given in (\ref{modCSresidual}).
Simulation was done according to the model specified above. We used random Gaussian measurements in this simulation, i.e. we generated $A$ as an $n \times m$ matrix with iid zero mean Gaussian entries and normalized each column to unit $\ell_2$ norm.

We experimented with two choices of $n$, $n=0.13m$ (where reg-mod-BPDN outperforms mod-BPDN) and $n=0.3m$ (where both are similar) and two values of $\sigma_p^2$, $\sigma_p^2=0.001$ (good prior) and $\sigma_p^2=0.1$ (bad prior). For the cases of Fig \ref{reconfig}(a) ($n=0.13m$, $\sigma_p^2=0.001$)  and Fig \ref{reconfig}(b) ($n=0.13m$, $\sigma_p^2=0.1$), we used signal length $m=256$, support size $|N|=0.1m=26$ and support extras size, $|\Delta_e|=0.1|N|=3$. The misses' size, $|\Delta|$, was varied between 0 and $0.2|N|$ (these numbers were motivated by the medical imaging application, we used larger numbers than what are shown in Fig. \ref{slowchange}). We used $\beta_l=1$, $\beta_m=0.4$ and $\beta_s=0.2$. The noise variance was $\sigma_w^2=10^{-5}$.
For the last two figures, Fig \ref{reconfig}(c) ($n=0.3m$, $\sigma_p^2=0.001$) and Fig \ref{reconfig}(d) ($n=0.3m$, $\sigma_p^2=0.1$), for which $n$ was larger, we used $\beta_m= \beta_s = 0.25$ which is a more difficult case for reg-mod-BPDN.  For Fig. \ref{reconfig}(c), we also used a larger noise variance $\sigma_w^2=10^{-4}$. All other parameters were the same.

{ In Fig. \ref{fig_extend}, we show a plot of reg-mod-BPDN and BPDN from Fig \ref{fig3a} extended all the way to $|\Delta|/|N|=1$ (which is the same as $\Delta = N$). Notice that if $|\Delta_e|=0$, then the point $|\Delta|/|N|=1$ of reg-mod-BPDN (or of mod-BPDN) is the same as BPDN. But in this plot, $|\Delta_e|=3$ and hence the two points are different, even though the errors are quite similar.}

{For applications where some training data is available, $\gamma$ and $\lambda$ for reg-mod-BPDN can be chosen by interpreting the reg-mod-BPDN solution as the maximum a posteriori (MAP) estimate under a certain prior signal model} (assume $x_T$ is Gaussian with mean $\hat{\mu}_T$ and variance $\sigma_p^2$ and $x_{T^c}$ is independent of $x_T$ and is iid Laplacian with parameter $b$). This idea is explained in detail in \cite{modcsjournal}. However, there is no easy way to do this for the other methods. Alternatively, choosing $\gamma$ and $\lambda$ according to Theorem 3 gives another good start point. We can do this for mod-BPDN and BPDN, but we cannot do this for the other methods (we show examples using this approach later). 
%
For a fair error comparison, for each algorithm, we selected $\gamma$ from a set of values $[0.00001\ 0.00005\ 0.0001\ 0.0005\ 0.001\ 0.005\ 0.01\ 0.1]$. We tried all these values for a small number of simulations (10 simulations) and then picked the best one (one with the smallest N-RMSE) for each algorithm. For weighted $\ell_1$ reconstruction, we also pick the best $\gamma'$ in (\ref{weightedl1}) from the same set in the same way\footnote{To give an example, our finally selected numbers for Fig. \ref{reconfig}(d) were $\gamma=0.01,0.001, 0.001,0.001,0.001,0.001,0.01,0.01$ for BPDN, mod-BPDN, reg-mod-BPDN, weighted $\ell_1$, LS-CS, CS-residual, CS-mod-residual, mod-CS-residual respectively and  $\gamma'=0.0001$}. For reg-mod-BPDN, $\lambda$ should be larger when the signal estimate is good and should be decreased when the signal estimate is not so good. We can use $\lambda=\alpha \sigma_w^2 / \sigma_p^2$ to adaptively determine its value for different choices of $\sigma_w^2$ and $\sigma_p^2$. In our simulations, we used $\alpha = 0.2$ for Fig. \ref{reconfig} (a), (b) and (d) and $\alpha = 0.05$ for Fig. \ref{reconfig}(c).




We fixed the chosen $\gamma$, $\gamma'$ and $\lambda$ and did Monte Carlo averaging over 100 simulations. We conclude the following. (1) When the signal estimate is not good (Fig. \ref{reconfig}(b),(d)) or when $n$ is small (Fig. \ref{reconfig}(a),(b)), CS-residual and CS-mod-residual have significantly larger error than reg-mod-BPDN. (2) In case of Fig. \ref{reconfig}(d) ($n=0.3m$), they also have larger error than mod-BPDN.
%
(3) In all four cases, weighed $\ell_1$ and mod-BPDN have similar performance. This is also similar to that of reg-mod-BPDN in case of  $n=0.3m$, but is much worse in case of $n=0.13m$.
(4) We also show a comparison with regmodBPDN-var in Fig. \ref{reconfig}(a). Notice that it has larger errors than reg-mod-BPDN for reasons explained in Sec. I-C.


\begin{figure}
\center
\includegraphics [width=9cm,height=6cm]{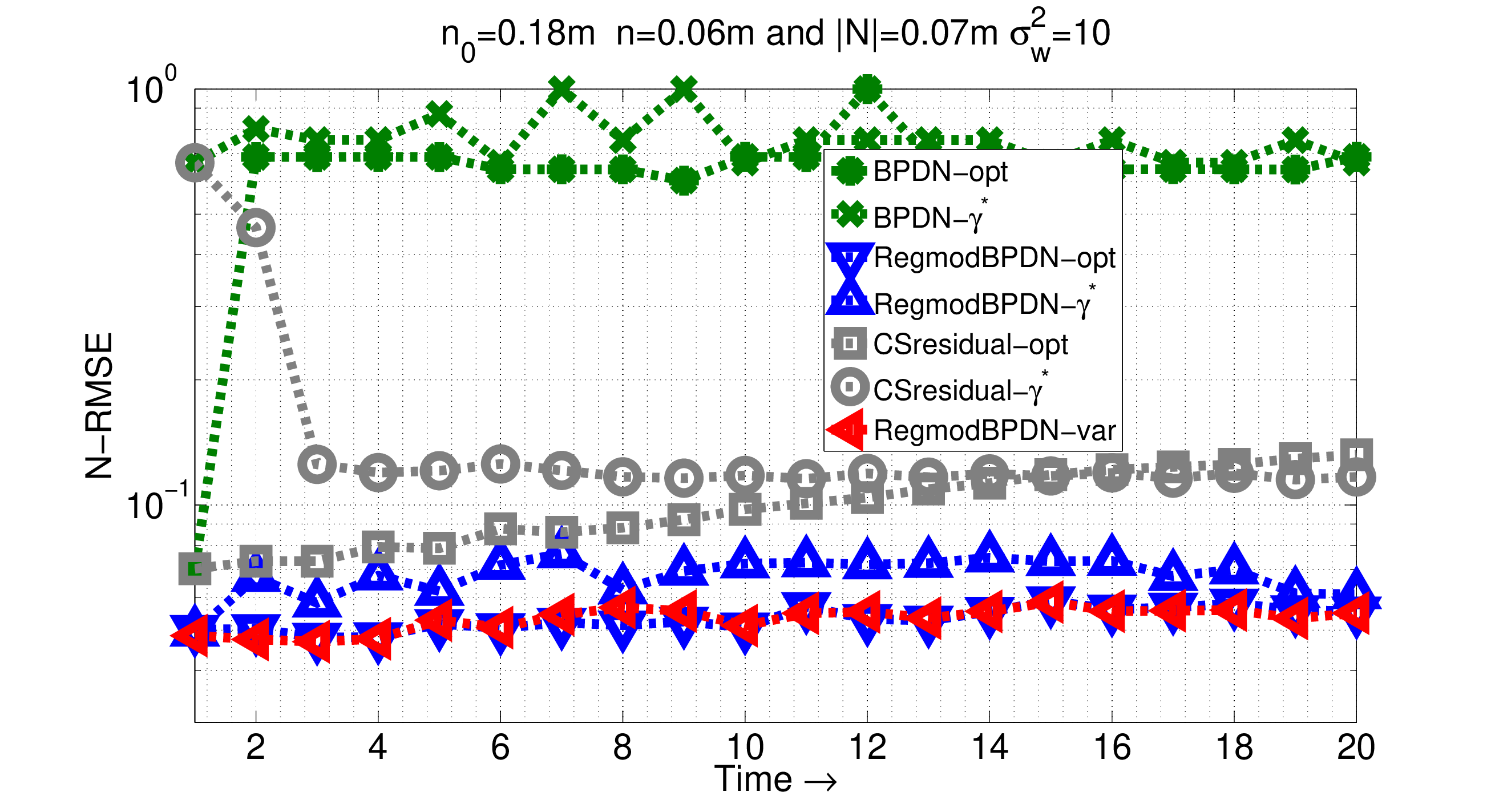}
\vspace{-4mm}
\caption{\small{Reconstructing a $32 \times 32$ block of the actual (compressible) larynx sequence from partial Fourier measurements.
Measurements $n=0.18m$ for $t=0$ and $n=0.06m$ for $t>0$.
Reg-mod-BPDN has the smallest reconstruction error among all methods.}
} \label{larynxcompareregmodBPDN}
\end{figure}

\subsection{Dynamic MRI application using $\gamma$ from Theorem 3}
In Fig. \ref{larynxcompareregmodBPDN}, we show comparisons for simulated dynamic MR imaging of an actual larynx image
sequence (Fig. \ref{slowchange} (a)(i)). The larynx image is not exactly sparse but is only compressible in the wavelet domain.
We used a two-level Daubechies-4 2D discrete wavelet transform (DWT). The $99$\%-energy support size of its wavelet transform vector,
$|N_t|\approx 0.07m$. Also, $|\Delta_t|\approx 0.001m$ and $|\Delta_{e,t}|\approx 0.002m$. We used a $32\times 32$ block of this sequence and at
each time and simulated undersampled MRI, i.e. we selected $n$ 2D discrete Fourier transform (DFT) coefficients using the variable density
sampling scheme of \cite{sparseMRI},  and added iid Gaussian noise with zero mean and variance $\sigma_w^2=10$ to each of them. Using a small $32\times 32$ block allows easy implementation using CVX (for full sized image sequences, one needs specialized code). We used $n_0=0.18m$ at $t=0$ and $n=0.06m$ at $t>0$.

We implemented dynamic reg-mod-BPDN as described in Algorithm 1. In this problem, the matrix $A=F_{u}\cdot W^{-1}$ where $F_u$ contains the selected rows of the 2D-DFT matrix and $W$ is the inverse 2D-DWT matrix (for a two-level Daubechies-4 wavelet). Reg-mod-BPDN was compared with similarly implemented reg-mod-BPDN-var and CS-residual algorithms (CS-residual only solved simple BPDN at $t=0$). We also compared with simple BPDN (BPDN done for each frame separately). For reg-mod-BPDN and reg-mod-BPDN-var, the support estimation threshold, $\rho$, was chosen as suggested in \cite{modcsjournal}: we used $\rho=20$ which is slightly larger than the smallest magnitude element in the $99\%$-energy support which is $15$.  At $t=0$, we used $T_0$ to be the set of indices of the wavelet approximation coefficients.
To choose $\gamma$ and $\lambda$ we tried two different things. (a)  We used $\lambda$ and $\gamma$ from the set
$[0.00001\ 0.00005\ 0.0001\ 0.0005\ 0.001\ 0.005\ 0.01\ 0.1]$ to do the reconstruction for a short training sequence (5 frames),
and used the average error to pick the best $\lambda$ and $\gamma$. We call the resulting reconstruction error plot reg-mod-BPDN-opt.
(b) We computed the average of the $\gamma^*$ obtained from Theorem 3 for the 5-frame training sequence and used this as $\gamma$ for the test sequence.
We selected $\lambda$ from the above set by choosing the one that minimizes the average of the bound of Theorem 3 for the 5 frames.
We call the resulting error plot reg-mod-BPDN-$\gamma^*$. The same two things were also done for BPDN and CS-residual as well. For reg-mod-BPDN-var,
we only did (a).

From Fig. \ref{larynxcompareregmodBPDN}, we can conclude the following. (1) Reg-mod-BPDN significantly outperforms the other methods when using so few measurements.  (2) Reg-mod-BPDN-var and reg-mod-BPDN have similar performance in this case. (3) The reconstruction performance of reg-mod-BPDN using $\gamma^*$ from Theorem 3 is close to that of reg-mod-BPDN using the best $\gamma$ chosen from a large set. This indicates that Theorem 3 provides a good way to select $\gamma$ in practice.

\subsection{Comparing the result of Theorem 1}
In Table I, we compare the result of Theorem 1 for reg-mod-BPDN, mod-BPDN and BPDN.
We used $m=256$, $|N|=26=0.1m$, $|\Delta|=0.04|N| = |\Delta_e|$, $\sigma_p^2=10^{-3}$, $\beta_l=1$ and $\beta_m=\beta_s=0.25$. Also, $\sigma_w^2=10^{-5}$ and we varied $n$. For each experiment with a given $n$, we did the following. We did $100$ Monte Carlo simulations. Each time, we evaluated the sufficient conditions for the bound of reg-mod-BPDN  to hold.  We say the bound {\em holds} if all the sufficient conditions hold for at least $98$ realizations. If this did not happen, we record {\em not hold} in Table I. If this did happen, then we recorded $\sqrt{\frac{\E[\text{bound}^2]}{\E[\|x\|^2_2]}}$ where $\E[\cdot ]$ denotes the Monte Carlo average computed over those realizations for which the sufficient conditions do hold. Here, ``bound" refers to the right hand side of (\ref{loosebound}) computed with $\gamma = \gamma^*_{T,\lambda}(\Delta)$ given in (\ref{regmodBPDNcond}). An analogous procedure was followed for both mod-BPDN and BPDN.

The comparisons are summarized in Table I. For reg-mod-BPDN, we selected $\lambda$ from the set $[0.00001 \ 0.00005 \ 0.0001 \ 0.0005 \ 0.001 \ 0.005 \ 0.01 \  0.1]$ by picking the one that gave the smallest bound. Clearly the reg-mod-BPDN result holds with the smallest $n$, while the BPDN result needs a very large $n$ ($n \ge 90\%$). Also even with $n=90\%$, the BPDN error bound is very large.
\begin{table}[h]
\center
\begin{tabular}{|c|c|c|c|}
  \hline
  $n$ & Reg-mod-BPDN & Mod-BPDN & BPDN \\
  \hline
  $0.13m$ & $0.885$ & not hold &  not hold \\
  \hline
  $0.19m$ & $0.161$ & $0.303$ &  not hold \\
  \hline
  $0.5m$ & $0.0199$ & $0.0199$ & not hold \\
  \hline
  $0.9m$ & $0.014$ & $0.014$ & $0.27$ \\
  \hline
\end{tabular}\label{tableb}
\caption{\small{Sufficient conditions and normalized bounds comparison of reg-mod-BPDN, mod-BPDN and
BPDN. Signal length $m=256$, support size $|N|=0.1m$, $|\Delta|=4\%|N|$, $\Delta_e=4\%|N|$, $\sigma_w^2=10^{-5}$ and $\sigma_p^2=10^{-3}$. ``not hold" means the one or all of the sufficient conditions does not hold.}}
\end{table}

\subsection{Comparing Theorems 1, 2, 3} 
\begin{figure*}[t!]
\centerline{
\hspace{-8mm}
\subfigure[$n=0.13m$,$\sigma_p^2=10^{-3}$,$\sigma_w^2=10^{-5}$]{
\includegraphics [width=6.1cm,height=4.8cm]{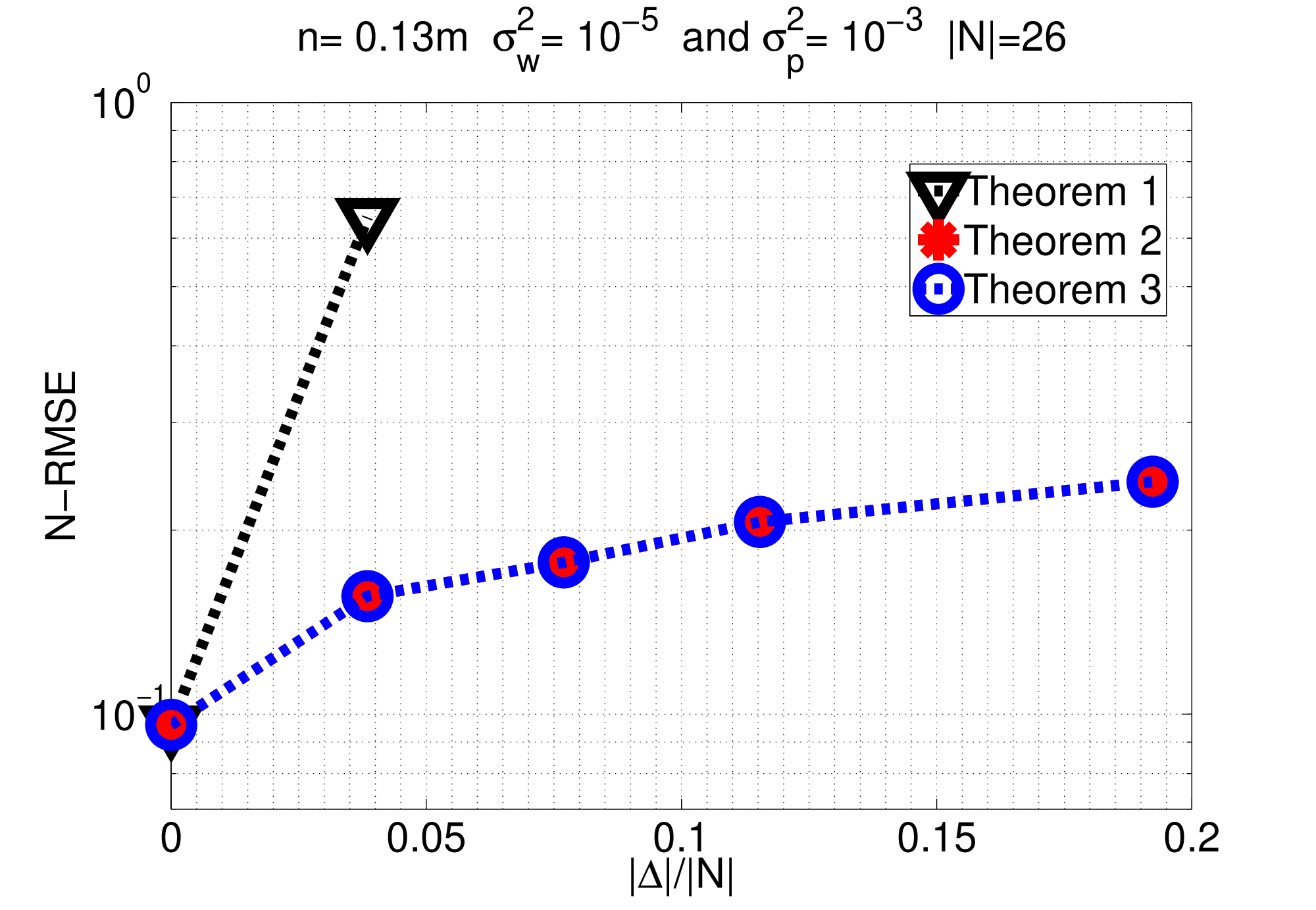}
}
\hspace{-8mm}
\subfigure[$n=0.13m$,$\sigma_p^2=10^{-3}$,$\sigma_w^2=10^{-5}$]{
\includegraphics [width=6.1cm,height=4.8cm]{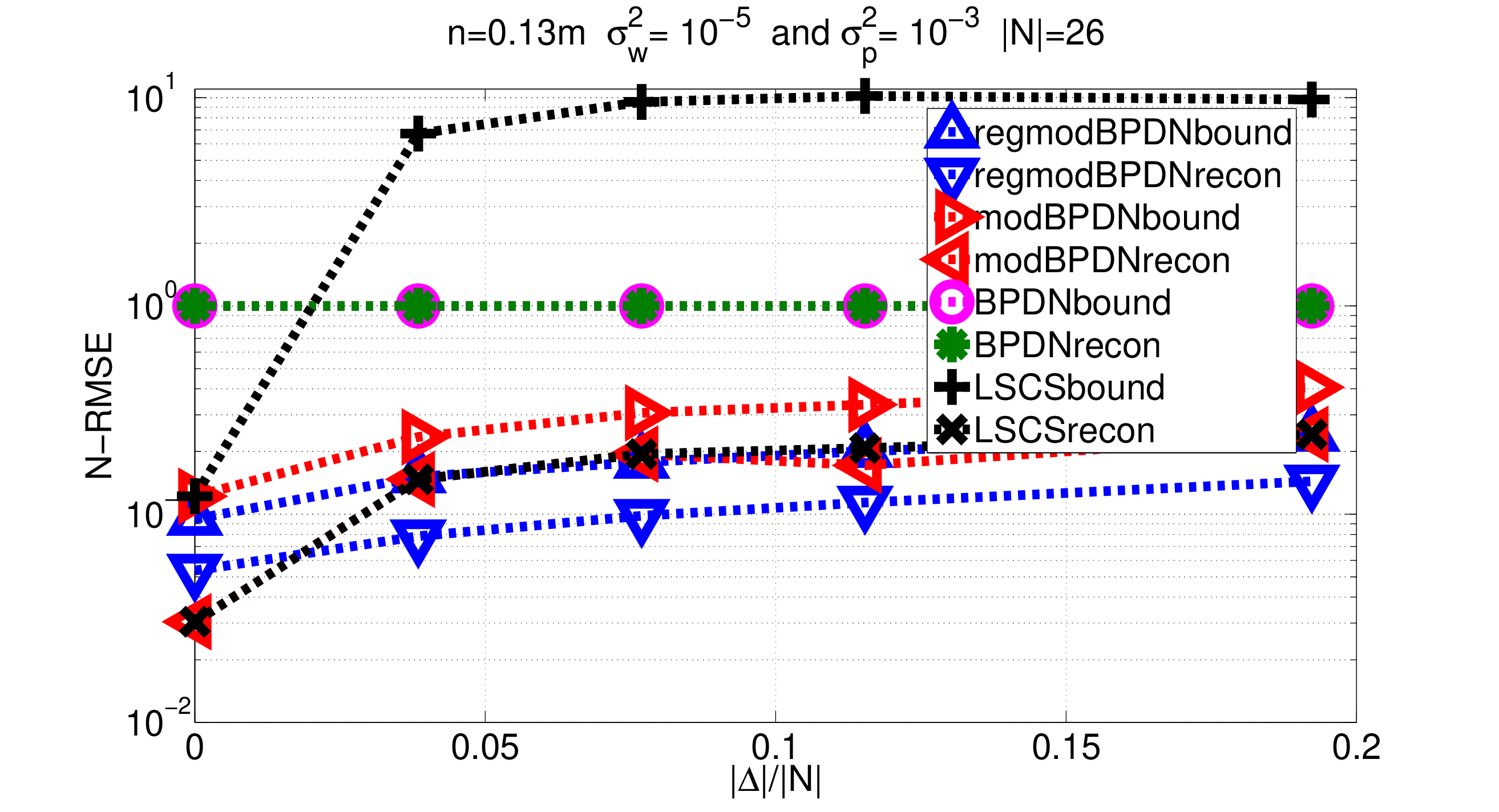}
}
\hspace{-8mm}
\subfigure[$n=0.17m$,$\sigma_p^2=10^{-3}$,$\sigma_w^2=10^{-3}$]{
\includegraphics [width=6.1cm,height=4.8cm]{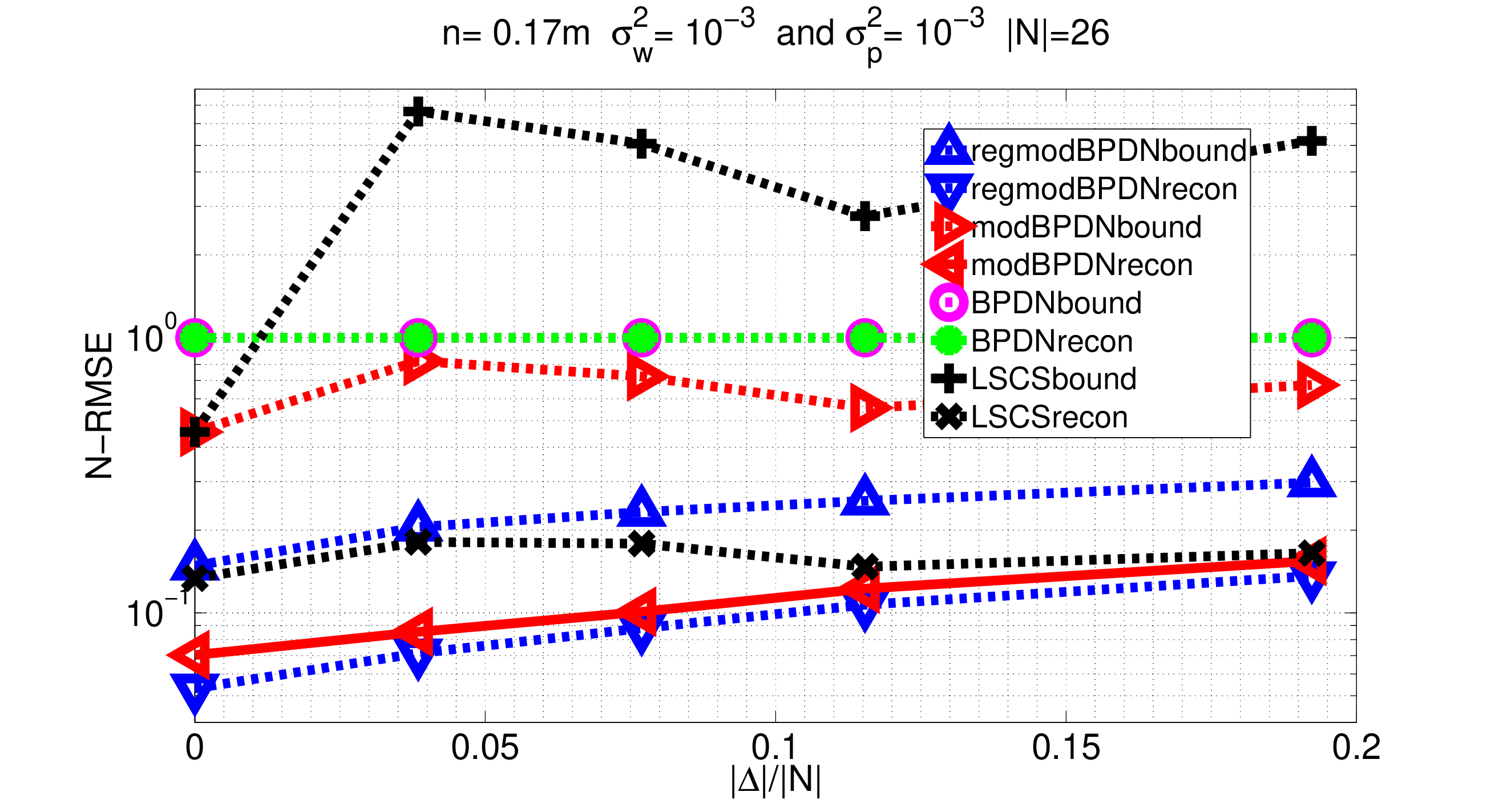}
}
}
\vspace{-1mm}
\caption{\small{In (a), we compare the three bounds from Theorem 1, 2 and 3  for one realization of $x$. In (b) and (c), we compare the normalized average bounds from Theorem 3 and reconstruction errors with random Gaussian and partial Fourier measurements respectively.}
} \label{bound4comparefig}
\end{figure*}

In Fig. \ref{bound4comparefig} (a), we compare the results from Theorems 1, 2 and 3 for one simulation. We plot $\frac{\text{bound}}{\|x\|_2}$ for $|\Delta|/|N|$ ranging from 0 to 0.2. Also, we used $m=256$, $|N|=26$, $|\Delta_e|=0.1|N|$, $\sigma_p^2=10^{-3}$, $\beta_l=1$ and $\beta_m=\beta_s=0.25$. Also, $n=0.13m$ and $\sigma_w^2=10^{-5}$. We used $\gamma = \gamma^*$ given in the respective theorems, and we set $\lambda= 10 \sigma_w^2 / \sigma_p^2$.
%
We notice the following. (1) The bound of Theorem 1 is much larger than that of Theorem 2 or 3, even for $|\Delta|=0.04|N|$. (2) For larger values of $|\Delta|$, the sufficient conditions of Theorem 1 do not hold and hence it does not provide a bound at all. (3) For reasons explained in Sec. IV, in this case, the bound of Theorem 3 is equal to that of Theorem 2. Recall that the computational complexity of the bound from Theorem 2 is exponential in $|\Delta|$. However if $|\Delta|$ is small, e.g. in our simulations $|\Delta| \le 5$, this is still doable.

\subsection{Upper bound comparisons using Theorem 3}
In Fig. \ref{bound4comparefig}(b), we do two things. (1) We compare the reconstruction error bounds from Theorem 3 for reg-mod-BPDN, mod-BPDN and BPDN and compare them with the bounds for LS-CS error given in \cite[Corollary 1]{LSCSbound}. All bounds hold without any sufficient conditions which is what makes this comparison possible.
(2) We also use the $\gamma^*$ given by Theorem 3 to obtain the reconstructions and compute the Monte Carlo averaged N-RMSE. Comparing this with the Monte Carlo averaged upper bound on the N-RMSE,$\sqrt{\frac{\E[\text{bound}^2]}{\E[\|x\|^2_2]}}$, allows us to evaluate the tightness of a bound. Here $\E[\cdot]$ denotes the mean computed over 100 Monte Carlo simulations and ``bound" refers to the right hand side of (\ref{regmodBPDNmediumtightbound}).
%
We used $m=256$, $|N|=26$, $|\Delta_e|=0.1|N|$, $\sigma_p^2=10^{-3}$, $\beta_l=1$, $\beta_m = \beta_s=0.25$, and $|\Delta|$ was varied from 0 to $0.2|N|$. Also, $n=0.13m$ and $\sigma_w^2=10^{-5}$.

From the figure, we can observe the following. (1) Reg-mod-BPDN has much smaller bounds than those of mod-BPDN, BPDN and LS-CS. The differences between reg-mod-BPDN and mod-BPDN bounds is minor when $|\Delta|$ is small but increases as $|\Delta|$ increases. (2) The conclusions from the reconstruction error comparisons are similar to those seen from the bound comparisons, indicating that the bound can serve as a useful proxy to decide which algorithm to use when (notice bound computation is much faster than computing the reconstruction error). (3) Also, reg-mod-BPDN and mod-BPDN bounds are quite tight as compared to the LS-CS bound. BPDN bound and error are both $100\%$. 100\% error is seen because the reconstruction is the all zeros' vector.

%

In Fig. \ref{bound4comparefig}(c), we did a similar set of experiments for the case where $A$ corresponds to a simulated MRI experiment, i.e. $A=F_{u}\cdot W^{-1}$ where $F_u$ contains randomly selected rows of the 2D-DFT matrix and $W$ is the inverse 2D-DWT matrix (for a two-level Daubechies-4 wavelet). We used $n=0.17m$ and $\sigma_w^2=10^{-3}$. All other parameters were the same as in Fig. \ref{bound4comparefig}(b). Our conclusions are also the same.


The complexity for Theorem 3 is polynomial in $|\Delta|$ whereas that of the LS-CS bound \cite[Corollary 1]{LSCSbound} is exponential in $|\Delta|$. To also show comparison with the LS-CS bound,
we had to choose a small value of $m=256$ so that the maximum value of $|\Delta|= 0.2|N|=5$ was small enough. In terms of MATLAB time, computation of the Theorem 3 bound for reg-mod-BPDN took 0.2 seconds while computing the LS-CS bound took 1.2 seconds. For all methods except LS-CS, we were able to do the same thing fairly quickly even for $m=4096$, or even larger. It took only $8$ seconds to compute the bound of Theorem 3 when $m=4096$, $n=0.13m$, $|N|=410=0.1m$ and $|\Delta|=|\Delta_e|=0.1|N|=41$.


{{

\section{Conclusions and Future Work}
In this work we studied the problem of sparse reconstruction from noisy undersampled measurements when partial and partly erroneous, knowledge of the signal's support and an erroneous estimate of the signal values on the ``partly known support" is also available. Denote the support knowledge by $T$ and the signal value estimate on $T$ by $\hat{\mu}_T$. We proposed and studied a solution called regularized modified-BPDN which tries to find the signal that is sparsest outside $T$, while being ``close enough" to $\hat{\mu}_T$ on $T$, and while satisfying the data constraint. We showed how to obtain computable error bounds that hold without any sufficient conditions. This made it easy to compare bounds for the various approaches (corresponding results for modified-BPDN and BPDN follow as direct corollaries). Empirical error comparisons with these and many other existing approaches are also provided.

In ongoing work, we are evaluating the utility of reg-mod-BPDN for recursive functional MR imaging to detect brain activation patterns in response to stimuli \cite{icip11_fmri}. On the other end, we are also working on obtaining conditions under which it will remain ``stable" (its error will be bounded by a time-invariant and small value) for a recursive recovery problem. In \cite{regmodBPDNstability}, this has been done for the constrained version of reg-mod-BPDN. {That result uses the restricted isometry constants (RIC) and the restricted orthogonality constants (ROC) \cite{decodinglp,candes_rip} in its sufficient conditions and bounds. However, this means that the conditions and bounds are not computable. Also, since the stability holds under a different set of sufficient conditions and has a different error bound than that for mod-CS \cite{stability_allerton} or LS-CS \cite{LSCS} or CS \cite{candes_rip}, comparison of the various results is difficult. An open question is how to extend the results of the current work (which are computable) to show the stability of unconstrained reg-mod-BPDN.}

\appendix

\subsection{Proof of Proposition 1} \label{Appprop1}
When $\lambda=0$, $Q_{T,0}(S)={A_{T \cup S}}'A_{T \cup S}$. Thus, $Q_{T,\lambda}(S)$ is invertible iff $A_{T \cup S}$ is full rank. When $\lambda>0$, $Q_{T,\lambda}(S)$ is as defined in (\ref{def_Q}).
Apply block matrix inversion lemma
\begin{eqnarray}
& & \hspace{-1mm} \left[
  \begin{array}{cc}
    \mathbf{A}\quad & \mathbf{B} \\
    \mathbf{C}\quad & \mathbf{D} \\
  \end{array}
\right]^{-1}= \nn \\
& & \hspace{-1mm} \left[
  \begin{array}{cc}
    (\mathbf{A}-\mathbf{BD}^{-1}\mathbf{C})^{-1} \ \ & -(\mathbf{A}-\mathbf{BD}^{-1}\mathbf{C})^{-1}\mathbf{BD}^{-1}  \\
    -\mathbf{D}^{-1}\mathbf{C}(\mathbf{A}-\mathbf{BD}^{-1}\mathbf{C})^{-1} \ \ & \mathbf{D}^{-1}+\mathbf{D}^{-1}\mathbf{C}(\mathbf{A}-\mathbf{BD}^{-1}\mathbf{C})^{-1}\mathbf{BD}^{-1} \\
  \end{array}
\right]\nn
\end{eqnarray}
with $\mathbf{A}={A_T}'A_T+\lambda I_T$, $\mathbf{B}={A_T}'A_S$, $\mathbf{C}={A_S}'A_T$ and $\mathbf{D}={A_S}'A_S$,
clearly $Q_{T,\lambda}(S)$ is invertible iff ${A_S}'A_S$ and ${A_T}'R A_T + \lambda I_T$ are invertible where $R:=[I-A_S({A_{S}}'A_{S})^{-1}A_{S}']$. When $A_S$ is full rank, (i) ${A_S}'A_S$ is full rank; and (ii) $R$ is a projection matrix. Thus $R=R'R$ and so ${A_T}'R A_T = (R A_T)' (R A_T)$ is positive semi-definite. As a result,
${A_T}'R A_T + \lambda I_T$ is positive definite and thus invertible. Hence, when $A_S$ is invertible, $Q_{T,\lambda}(S)$ is also invertible.

\subsection{Proof of Theorem 1} \label{App3lemmas}
In this subsection, we give the three lemmas for the proof of Theorem 1. {\em To keep notation simple we remove the subscripts $_{T,\lambda}$ from $Q(\Delta)$, $M$, $P(\Delta)$, $d(\Delta)$, $c(\Delta)$, $ERC(\Delta)$ in this and other Appendices.}
\begin{lemma}\label{lm1}
Suppose that $Q({\Delta})$ is invertible, then
\begin{equation}
 \|d(\Delta)-c(\Delta)\|_2\le \gamma \sqrt{|\Delta|}\cdot f_1(\Delta) \label{step1eq}
\end{equation}
\end{lemma}

Lemma 1 can be obtained by setting $\nabla L(b)=0$ and then using block matrix inversion on $Q(\Delta)$. The proof of Lemma 1 is in Appendix \ref{Applemma1proof}. Next, $\|c(\Delta)-x\|_2$ can be bounded using the following lemma.

\begin{lemma}
Suppose that $Q(\Delta)$ is invertible. Then
\begin{eqnarray}
\hspace{-5mm}\|c(\Delta)-x\|_2 \le \lambda f_2(\Delta) \|x_{T}-\hat{\mu}_{T}\|_2+f_3(\Delta)\|w\|_2 \label{regmodBPDNbound}
\end{eqnarray}

\end{lemma}
The proof of Lemma 2 is in Appendix \ref{Applemma2proof}.

\begin{lemma}
If $Q({\Delta})$ is invertible, $ERC(\Delta)>0$, and $\gamma\ge \gamma^*(\Delta)$, then $L(b)$ has a unique minimizer which is equal to $d(\Delta)$ .
\end{lemma}

Lemma 3 can be obtained in a fashion similar to \cite{justrelax,modBPDN}.  Its proof is given in Appendix \ref{Applemma3proof}.

Combining Lemmas 1, 2 and 3, and using the fact $\|d(\Delta)-x\|_2\le \|d(\Delta)-c(\Delta)\|_2+\|c(\Delta)-x\|_2$, we get Theorem 1.

\subsection{Proof of Theorem 2} \label{Applemma4}
The following lemma is needed for the proof of the corollaries leading to Theorem 2.
\begin{lemma} \label{lemma4}
Suppose that $Q({\tDelta})$ is invertible. Then
\begin{eqnarray}
& &\|c(\tilde{\Delta})- x\|_2
\le \nn \\ & & \lambda  f_2(\tilde{\Delta}) \|x_{T}-\hat{\mu}_{T}\|_2
 +f_3(\tilde{\Delta})\|w\|_2+
  f_4(\tilde{\Delta}) \|x_{\Delta \setminus \tilde{\Delta}}\|_2 \ \ \label{lemma4bound}
\end{eqnarray}
\end{lemma}
Since $c(\tilde{\Delta})$ is only supported on $T\cup \tilde{\Delta}$ and $y=A_{T\cup \tilde{\Delta}}x_{T\cup \tilde{\Delta}}+A_{\Delta \setminus \tilde{\Delta}}x_{\Delta \setminus \tilde{\Delta}}+w$, the last term of (\ref{lemma4bound}) can be obtained by separating $x_{\Delta \setminus \tilde{\Delta}}$ out. The proof of Lemma \ref{lemma4} is given in Appendix \ref{Applemma4proof}.

Using Lemma 4, we can obtain Corollary 1 and then Corollary 2. Then minimize over all allowed $\tDelta$'s in Corollary 2, we get Theorem 2. The proof of Corollary 1 and 2 are given as follows.

\subsubsection{Proof of Corollary 1}\label{Appcor1proof}
Notice from the proof of Lemma 1 and Lemma 3 that nothing in the result changes if we replace $\Delta$ by a $\tDelta \subseteq \Delta$.
By Lemma 1 for $\tDelta$, we are able to bound $\|d(\tDelta)-c(\tDelta)\|_2$. Hence, we get the first term of (\ref{def_f}).
Next, invoke Lemma 4 to bound $\|c(\tDelta)-x\|_2$ and we can obtain the rest three terms of (\ref{def_f}). Lemma 3 for
$\tDelta$ gives the sufficient conditions under which $d(\tDelta)$ is the unique unconstrained minimizer of $L(b)$.

\subsubsection{Proof of Corollary 2}\label{Appcor2proof}
Corollary 2 is obtained by bounding $\gamma^*(\tilde{\Delta})$.
$\gamma^*(\tilde{\Delta}) = \|{A_{(T\cup \tDelta)^c}}'(y - A c(\tDelta))\|_{\infty} / ERC(\tDelta)$ can be bounded by
rewriting  $y-Ac(\tilde{\Delta}) = A_{T\cup \Delta}(x_{T\cup \Delta}-(c(\tilde{\Delta}))_{T\cup \Delta} ) + w$
and then bounding $\|x_{T\cup \Delta}-(c(\tilde{\Delta}))_{T\cup \Delta}\|_2 = \|x-c(\tilde{\Delta})\|_2$ using Lemma 4. Doing this, we get
\begin{eqnarray}
& &\|{A_{(T\cup \tDelta)^c}}'(y - A c(\tDelta))\|_{\infty}\nn \\
& & \le  \max_{i \notin T \cup \tDelta}\ \  |{A_i}'A_{T \cup \Delta}(x_{T\cup \Delta}-(c(\tilde{\Delta}))_{T\cup \Delta})|+|{A_i}'w|\nn \\
& & \le  \max_{i \notin T \cup \tDelta}\ \  \|{A_i}'A_{T \cup \Delta}\|_2 \|x_{T\cup \Delta}-(c(\tilde{\Delta}))_{T\cup \Delta})\|_2+|{A_i}'w|\nn \\
& &  \le \text{maxcor}(\tDelta) \lambda f_2(\tDelta)\|x_{T}-\mu_T\|_2+ \text{maxcor}(\tDelta) f_3(\tDelta)\|w\|_2 \nn \\
& & +\text{maxcor}(\tDelta)f_4(\tDelta)\|x_{\Delta \setminus \tDelta}\|_2 +\|{A_{(T\cup \tDelta)^c}}'w\|_{\infty} \nn
\end{eqnarray}
Using the above inequality to bound $\gamma^*(\tDelta)$ and
replacing $\gamma$ in $f(T,\lambda,\Delta,\tilde{\Delta},\gamma)$, given in (\ref{def_f}), by this bound, we can get (\ref{gdefbound}).

\subsection{Proof of Lemmas 1, 2, 3, 4}
\subsubsection{Proof of Lemma 1} \label{Applemma1proof}
We use the approach of \cite[Lemma 3]{justrelax}. We can minimize the function $L(b)$
over all vectors supported on set $T\cup \Delta$ by minimizing:
\begin{equation}
F(b)=\frac{1}{2}\|y-A_{T\cup \Delta}b_{T\cup \Delta}\|_2^2+\frac{1}{2}\lambda\|b_T-\hat{\mu}_T\|_2^2+\gamma\|b_{\Delta}\|_1
\end{equation}
Since $Q(\Delta)$ is invertible, $F(b)$ is strictly convex as a function of $b_{T\cup \Delta}$. Then at the unique minimizer, $d(\Delta)$, $0\in \nabla F(b)|_{b=d(\Delta)}$. Let $\partial \|b_{T^c}\|_1|_{b=d(\Delta)}$ denote
the subgradient set of $\|b_{T^c}\|_1$ at $b=d(\Delta)$. Then clearly any $\phi$ in this set satisfies
\begin{eqnarray}
\phi_T &=& 0 \label{phiT} \\
\|\phi_{T^c}\|_{\infty} &\le & 1
\end{eqnarray}
Now, $0\in \nabla F(b)|_{b=d(\Delta)}$ implies that
\begin{eqnarray}
({A_{T\cup \Delta}}'A_{T\cup \Delta})[d(\Delta)]_{T\cup \Delta}-{A_{T\cup \Delta}}'y\nonumber\\+\lambda\left[
                                                          \begin{array}{c}
                                                            [d(\Delta)]_{T}-\hat{\mu}_T \\
                                                            \textbf{0}_{\Delta} \\
                                                          \end{array}
                                                        \right]
                                                        +\gamma \phi_{T\cup \Delta}=0
\end{eqnarray}
Simplifying the above equation, we get
\begin{equation}
[d(\Delta)]_{T\cup \Delta} =Q(\Delta)
^{-1}({A_{T\cup \Delta}}'y+\lambda \left[
                                      \begin{array}{c}
                                        \hat{\mu}_T \\
                                        \mathbf{0}_{\Delta} \\
                                      \end{array}
                                    \right]
-\gamma \phi_{T\cup \Delta})
\end{equation}
Therefore, using (\ref{phiT}) and (\ref{reglssolution}), we have
\begin{equation}
[c(\Delta)]_{T\cup \Delta}-[d(\Delta)]_{T\cup \Delta}=Q(\Delta)^{-1}\left[
                                                \begin{array}{c}
                                                  \textbf{0}_T \\
                                                  \gamma \phi_{\Delta} \\
                                                \end{array}
                                              \right]
\end{equation}
Since
\begin{equation}
Q(\Delta)= \left [ \begin{array}{cc}
{A_T}'A_T+\lambda I_{T}\ & {A_T}'A_{\Delta}\\
{A_{\Delta}}'A_T \ & {A_{\Delta}}'A_{\Delta}
\end{array} \right ],
\end{equation}
using the block matrix inversion lemma
\begin{eqnarray}
& & \hspace{-1mm}\left[
  \begin{array}{cc}
    \mathbf{A}\quad & \mathbf{B} \\
    \mathbf{C}\quad & \mathbf{D} \\
  \end{array}
\right]^{-1}=\nn \\
& & \hspace{-1mm} \left[
  \begin{array}{cc}
    \mathbf{A}^{-1}+\mathbf{A}^{-1}\mathbf{B}(\mathbf{D}-\mathbf{CA}^{-1}\mathbf{B})^{-1}\mathbf{CA}^{-1}\ & -\mathbf{A}^{-1}\mathbf{B}(\mathbf{D}-\mathbf{CA}^{-1}\mathbf{B})^{-1} \\
    -(\mathbf{D}-\mathbf{CA}^{-1}\mathbf{B})^{-1}\mathbf{CA}^{-1} \ & (\mathbf{D}-\mathbf{CA}^{-1}\mathbf{B})^{-1} \\
  \end{array}
\right] \nn
\end{eqnarray}
with $\mathbf{A}={A_T}'A_T+\lambda I_{T}$, $\mathbf{B}={A_T}'A_{\Delta}$, $\mathbf{C}={A_{\Delta}}'A_T$ and $\mathbf{D}={A_{\Delta}}'A_{\Delta}$ and using $\phi_T=0$, we obtain
\begin{eqnarray}
& &[c(\Delta)]_{T \cup \Delta}-[d(\Delta)]_{T \cup \Delta}=\nonumber\\
& & \left [ \begin{array}{c}
-\gamma ({A_T}'A_T+\lambda I_{|T|})^{-1}{A_T}A_{\Delta}({A_{\Delta}}'MA_{\Delta})^{-1}\phi_{\Delta}\\
\gamma ({A_{\Delta}}'MA_{\Delta})^{-1}\phi_{\Delta}
\end{array} \right ]\nonumber
\end{eqnarray}
Since $\|\phi_{\Delta}\|_{\infty} \le 1$, the bound of (\ref{step1eq}) follows.

\subsubsection{Proof of Lemma 2}\label{Applemma2proof}
Recall $c(\Delta)$ is given in (\ref{reglssolution}). Since both $x$ and $c(\Delta)$ are zero outside $T\cup \Delta$, then $\|c(\Delta)-x\|_2=\|[c(\Delta)]_{T\cup \Delta}-x_{T\cup \Delta}\|_2$. With $y=Ax+w$ and $Ax=A_{T\cup \Delta}x_{T\cup \Delta}$, we have
\begin{equation}
{A_{T\cup \Delta}}'y={A_{T\cup \Delta}}'(A_{T\cup \Delta}x_{T\cup \Delta}+w) \label{Ay}
\end{equation}
Notice ${A_{T\cup \Delta}'}A_{T\cup \Delta}=Q(\Delta)-\lambda \left[
                                                 \begin{array}{cc}
                                                   I_{T}\ & \mathbf{0}_{T,S} \\
                                                   \mathbf{0}_{S,T}\ & \mathbf{0}_{S,S} \\
                                                 \end{array}
                                               \right]$.
Using (\ref{Ay}), we obtain the following equation
\begin{eqnarray}
{A_{T\cup \Delta}}'y=
                                               Q(\Delta)x_{T\cup \Delta}-\lambda  \left[
                                                 \begin{array}{c}
                                                   x_{T}  \\
                                                   \mathbf{0}_{\Delta}  \\
                                                 \end{array}
                                               \right]+{A_{T\cup \Delta}}'w
\end{eqnarray}
Then, using (\ref{reglssolution}) we can obtain
\begin{eqnarray}
\hspace{0mm}[c(\Delta)]_{T\cup \Delta}-x_{T\cup \Delta}= \lambda Q(\Delta)^{-1}\left[
                                                                                        \begin{array}{c}
                                                                                          \hat{\mu}_{T}-x_T \\
                                                                                          \mathbf{0}_{\Delta} \\
                                                                                        \end{array}
                                                                                      \right]
+Q(\Delta)^{-1}{A_{T\cup \Delta}}'w \nn
\end{eqnarray}
Finally, this gives (\ref{regmodBPDNbound}).


\subsubsection{Proof of Lemma 3}\label{Applemma3proof}
The proof is similar to that in \cite{justrelax} and \cite{modBPDN}. Recall that $d(\Delta)$ minimizes the function $L(b)$ over all $b$ supported on $T\cup \Delta$. We need to show that if $\gamma \ge \gamma^*(\Delta)$, then $d(\Delta)$ is the unique global minimizer of $L(b)$.

The idea is to prove under the given condition, any small perturbation $h$ on
$d(\Delta)$ will increase function $L(d(\Delta))$,i.e.
$L(d(\Delta)+h)-L(d(\Delta))>0,\forall \|h\|_{\infty} \le \epsilon$ for $\epsilon$ small enough. Then since $L(b)$ is a convex function, $d(\Delta)$ will be the unique global minimizer\cite{justrelax}.

Similar to \cite{modBPDN}, we first split the perturbation into two parts $h=u+v$ where $u$ is supported on
$T\cup \Delta$ and $v$ is supported on $(T\cup \Delta)^c$. Clearly $\|u\|_{\infty}\le \|h\|_{\infty}\le \epsilon$. We consider the case $v \neq 0$ since the case $v=0$ is already covered in Lemma 1.  Then
\begin{eqnarray}
L(d(\Delta)+h)=\frac{1}{2}\|y-A(d(\Delta)+u)-Av\|_2^2+ \nonumber\\
\frac{1}{2}\lambda\|[d(\Delta)]_T+u_T+v_T-\hat{\mu}_T\|_2^2+
\gamma\|(d(\Delta)+u)_{T^c}+v_{T^c}\|_1 \nonumber
\end{eqnarray}

Then, we can obtain
\begin{eqnarray}
\hspace{-15mm}L(d(\Delta)+h)-L(d(\Delta))=
L(d(\Delta)+u)-L(d(\Delta))\nonumber\\
+\frac{1}{2}\|Av\|_2^2-\langle y-Ad(\Delta),Av \rangle + \langle Au,Av \rangle+\gamma \|v_{T^c}\|_1\nonumber
\end{eqnarray}
Since $d(\Delta)$ minimizes $L(b)$ over all vectors supported on $T\cup \Delta$, $L(d(\Delta)+u)-L(d(\Delta))\ge 0$. Then since $L(d(\Delta)+u)-L(d(\Delta))\ge 0$ and $\|Av\|_2^2 \ge 0$, we
need to prove that the rest are positive,i.e.,$\gamma\|v_{T^c}\|_1- \langle y-Ad(\Delta),Av\rangle +\langle Au,Av \rangle \ge
0$. Instead, we can prove this by proving a stronger condition
$\gamma\|v_{T^c}\|_1-|\langle y-Ad(\Delta),Av\rangle |-|\langle Au,Av \rangle |\ge 0$.
Since $\langle y-Ad(\Delta),Av\rangle =v'A'(y-Ad(\Delta))$ and $v$ is supported on $(T\cup \Delta)^c$,
\begin{eqnarray}
\hspace{-3mm}|\langle y-Ad(\Delta),Av\rangle |&=&|{v_{(T\cup \Delta)^c}}'{A_{(T\cup \Delta)^c}}'(y-Ad(\Delta))| \nonumber\\ &\le& \|v\|_1\|{A_{(T\cup \Delta)^c}}'(y-Ad(\Delta))\|_{\infty}\nonumber
\end{eqnarray}
Thus,
\begin{eqnarray}
|\langle y-Ad(\Delta),Av\rangle |
\le \max_{\omega \notin T\cup \Delta}|\langle y-Ad(\Delta),A_{\omega}\rangle \|v\|_1 \nn
\end{eqnarray}
Meanwhile,
\begin{equation}
|\langle Au,Av\rangle |\le \|A'Au\|_{\infty}\|v\|_1\le
\epsilon\|A'A\|_{\infty}\|v\|_1
\end{equation}
And $\|v\|_1=\|v_{T^c}\|_1$ since $v$ is supported on $(T\cup \Delta)^c \subseteq T^c$. Then what we need to prove is
\begin{equation}
\big{[}\gamma-\max_{\omega \notin
T\cup \Delta}|\langle y-Ad(\Delta),A_{\omega}\rangle |-\epsilon\|A'A\|_{\infty}\big{]}\|v\|_1 >0
\end{equation}
Since we can select $ \epsilon > 0$ as small as possible, then we just
need to show
\begin{equation}
\gamma-\max_{\omega \notin
T\cup \Delta}|\langle y-Ad(\Delta),A_{\omega}\rangle |>0
\end{equation}
Since $y-Ad(\Delta)=(y-Ac(\Delta))+A(c(\Delta)-d(\Delta))$, and by
Lemma 1 we know $A(c(\Delta)-d(\Delta))=\gamma MA_{\Delta}({A_{\Delta}}'MA_{\Delta})^{-1}\phi_{\Delta}$ and since $\|\phi_{\Delta}\|_{\infty} \le 1$, we conclude that $d(\Delta)$ is the unique global minimizer if
\begin{equation}
\hspace{-1mm}\|{A_{(T\cup \Delta)^c}}'(y-Ac(\Delta))\|_{\infty}< \gamma \big{[}1-\max_{\omega \notin
T\cup \Delta}\|P(\Delta){A_{\Delta}}'MA_{\omega}\|_1\big{]}\label{suffcondineq}
\end{equation}
Next, we will show that $d(\Delta)$ is also the unique global minimizer under the following condition
\begin{equation}
\hspace{-3mm}\|{A_{(T\cup \Delta)^c}}'(y-Ac_{T,\lambda}(\Delta))\|_{\infty}= \gamma \big{[}1-\max_{\omega \notin
T\cup \Delta}\|P(\Delta){A_{\Delta}}'MA_{\omega}\|_1\big{]}  \label{suffcondeq}
\end{equation}
Since the perturbation $h\neq 0$, then $u\neq 0$ or $v\neq 0$. Therefore, we will discuss the following three cases.
\begin{enumerate}
\item $u\neq 0$. In this case, we know $L(d(\Delta)+u)-L(d(\Delta))>0$ since $d(\Delta)$ is the unique minimizer over all vectors supported on $T\cup \Delta$. Therefore, $L(d(\Delta)+h)-L(d(\Delta))>0$ if (\ref{suffcondeq}) holds.

\item $u=0$, $v\neq 0$ and $v$ is not in the null space of $A$, i.e., $Av\neq 0$. In this case, we know $\|Av\|_2^2>0$. Hence, $L(d(\Delta)+h)-L(d(\Delta))>0$ when (\ref{suffcondeq}) holds.

\item $u=0$, $v\neq 0$ and $Av=0$. In this case, $L(d(\Delta)+h)-L(d(\Delta))=\gamma \|v_{T^c}\|_1$. Thus, $L(d(\Delta)+h)-L(d(\Delta))>0$ if $\gamma>0$. Clearly, $L(d(\Delta)+h)-L(d(\Delta))>0$ when (\ref{suffcondeq}) holds.
\end{enumerate}
Finally, combining (\ref{suffcondineq}) and (\ref{suffcondeq}),
we can conclude that $d(\Delta)$ is the unique global minimizer if the following condition holds
\begin{equation}
\|{A_{(T\cup \Delta)^c}}'(y-Ac(\Delta))\|_{\infty}\le \gamma \text{ERC}(\Delta)
\end{equation}

\subsubsection{Proof of Lemma 4} \label{Applemma4proof}
Consider a $\tilde{\Delta}\subseteq \Delta$ such that $A_{\tilde{\Delta}}$ has full rank.
Since ${A_{T\cup \tilde{\Delta}}}'y={A_{T\cup \tilde{\Delta}}}'(A_{T\cup \tilde{\Delta}}x_{T\cup \tilde{\Delta}}+w+A_{\Delta \setminus \tilde{\Delta}}x_{\Delta \setminus \tilde{\Delta}})$, expanding these terms we have
\begin{equation}
\hspace{-2mm}{A_{T\cup \tilde{\Delta}}}'y=Q(\Delta)x_{T\cup \tilde{\Delta}}-\lambda \left[
                                                 \begin{array}{c}
                                                   x_{T}  \\
                                                   \mathbf{0}_{\tilde{\Delta}}  \\
                                                 \end{array}\right]+{A_{T\cup \tilde{\Delta}}}'w+{A_{T\cup \tilde{\Delta}}}'A_{\Delta \setminus \tilde{\Delta}}x_{\Delta \setminus \tilde{\Delta}}
\end{equation}
Then, using this in the expression for $c(\tDelta)$ from (\ref{def_c}), we get
\begin{eqnarray}
& & [c(\tilde{\Delta})]_{T\cup \Delta}-x_{T\cup \Delta}=
 \left[ \begin{array}{c}
                                                                      \lambda Q(\tilde{\Delta})^{-1}\left[
                                                                                        \begin{array}{c}
                                                                                          \hat{\mu}_{T}-x_T \\
                                                                                          \mathbf{0}_{\tilde{\Delta}} \\
                                                                                        \end{array}
                                                                                      \right]
 \\
\mathbf{0}_{\Delta \setminus \tilde{\Delta}}\\
 \end{array}
 \right] \nonumber \\
 & & + \left[ \begin{array}{c}
 Q(\tilde{\Delta})^{-1}{A_{T\cup \tilde{\Delta}}}'w \\
\mathbf{0}_{\Delta \setminus \tilde{\Delta}}\\
 \end{array}
 \right]
 + \left[ \begin{array}{c}
 Q(\tilde{\Delta})^{-1}{A_{T\cup \tilde{\Delta}}}'A_{\Delta \setminus \tilde{\Delta}}x_{\Delta \setminus \tilde{\Delta}}\\
 -x_{\Delta \setminus \tilde{\Delta}}
  \\
                                                                                        \end{array}
                                                                                      \right]\ \
\quad \end{eqnarray}
Therefore, we get (\ref{lemma4bound}).

\bibliographystyle{IEEEtran}
\bibliography{regmodBPDNjournalbib,fMRImodcsbib}


\end{document}